\documentclass[10pt,conference]{IEEEtran}
\IEEEoverridecommandlockouts
\def\BibTeX{{\rm B\kern-.05em{\sc i\kern-.025em b}\kern-.08em
    T\kern-.1667em\lower.7ex\hbox{E}\kern-.125emX}}
    
\usepackage{cite}
\usepackage{placeins}
\usepackage{makecell}
\usepackage{amsmath,amssymb,amsfonts,amsthm,enumerate}
\usepackage{graphicx}
\usepackage{textcomp}
\usepackage{xcolor}
\usepackage{mathrsfs}
\usepackage{multirow}
\usepackage{array}
\usepackage{subcaption}
\usepackage{footnote}
\usepackage{paralist}
\usepackage{threeparttable}
\usepackage{booktabs}
\usepackage{color,framed}
\usepackage[table]{xcolor}
\usepackage{setspace}
\definecolor{shadecolor}{rgb}{0.92,0.92,0.92}
\definecolor{darkgreen}{RGB}{0,100,0}
\usepackage{tcolorbox}
\usepackage{caption}
\usepackage{url}
\usepackage{epstopdf}
\usepackage{balance}
\usepackage{enumitem}
\usepackage{tabularx}  
\usepackage{adjustbox}

\usepackage{array} 
\usepackage{booktabs} 
\usepackage{multirow} 
\usepackage[figuresright]{rotating}
\theoremstyle{definition}
\usepackage{pgfplots}

\newcommand{\benchmark}{{\sc ATSADBench}\xspace}

\newcommand{\Directly}{{\sc Direct}\xspace}
\newcommand{\Predictionbased}{{\sc Prediction-Based}\xspace}

\PassOptionsToPackage{hyphens}{url}
\expandafter\def\expandafter\UrlBreaks\expandafter{\UrlBreaks
  \do\a\do\b\do\c\do\d\do\e\do\f\do\g\do\h\do\i\do\j%
  \do\k\do\l\do\m\do\n\do\o\do\p\do\q\do\r\do\s\do\t%
  \do\u\do\v\do\w\do\x\do\y\do\z\do\A\do\B\do\C\do\D%
  \do\E\do\F\do\G\do\H\do\I\do\J\do\K\do\L\do\M\do\N%
  \do\O\do\P\do\Q\do\R\do\S\do\T\do\U\do\V\do\W\do\X%
  \do\Y\do\Z\do\*\do\-\do\~\do\'\do\"\do\-}%

\usepackage{hyperref}
\hypersetup{hidelinks,
	colorlinks=true,
	allcolors=blue,
	pdfstartview=Fit,
	breaklinks=true}
\usepackage{breakurl}
\usepackage{cleveref}
\crefname{section}{§}{§§}
\crefname{paragraph}{§}{§§}
\Crefname{figure}{Fig.}{Figs.}
\Crefname{definition}{Def.}{Defs.}

\makeatletter
\newif\if@restonecol
\makeatother

\usepackage[linesnumbered,ruled,vlined,noend]{algorithm2e}
\usepackage{algpseudocode}

\newenvironment{packeditemize}{
\begin{list}{$\bullet$}{
\setlength{\labelwidth}{8pt}
\setlength{\itemsep}{0pt}
\setlength{\leftmargin}{\labelwidth}
\addtolength{\leftmargin}{\labelsep}
\setlength{\parindent}{0pt}
\setlength{\listparindent}{\parindent}
\setlength{\parsep}{0pt}
\setlength{\topsep}{3pt}}}{\end{list}}

\usepackage[switch]{lineno}
\usepackage{pifont}
\tcbuselibrary{breakable}
\usepackage{circledsteps}

\usepackage{tikz}

\definecolor{OliveGreen}{rgb}{0,0.6,0}
\definecolor{BrickRed}{rgb}{0.8,0.2,0.2}
\definecolor{RoyalBlue}{rgb}{0.25, 0.41, 0.88}
\newcommand{\perfchange}[1]{%
  \ifdim#1pt>0.001pt \textcolor{OliveGreen}{+#1\%} \else \textcolor{BrickRed}{#1\%} \fi
}

\newcommand{\alchange}[1]{%
  \ifdim#1pt<0.001pt \textcolor{OliveGreen}{#1} \else \textcolor{BrickRed}{+#1} \fi
}

\begin{document}

\title{Evaluating Large Language Models for Time Series Anomaly Detection in Aerospace Software
\thanks{*Corresponding authors.}
}

\author{
\IEEEauthorblockN{
Yang Liu\textsuperscript{1},
Yixing Luo\textsuperscript{1}\textsuperscript{*},
Xiaofeng Li\textsuperscript{1},
Xiaogang Dong\textsuperscript{1}\textsuperscript{*},
Bin Gu\textsuperscript{1},
Zhi Jin\textsuperscript{2}
}

\IEEEauthorblockA{\textsuperscript{1}Software Department, Beijing Institute of Control Engineering, Beijing, China}
\IEEEauthorblockA{\textsuperscript{2}School of Computer Science, Peking University, Beijing, China}
\IEEEauthorblockA{
Email: liuyang12339@163.com,  \{luoyi\_xing, li\_x\_feng\}@126.com \\ dongxiaogang@263.net, gubinbj@sina.com, zhijin@pku.edu.cn}
}

\maketitle

\begin{abstract}
Time series anomaly detection (TSAD) is essential for ensuring the safety and reliability of aerospace software systems. Although large language models (LLMs) provide a promising training-free alternative to unsupervised approaches, their effectiveness in aerospace settings remains under-examined because of complex telemetry, misaligned evaluation metrics, and the absence of domain knowledge. To address this gap, we introduce \benchmark, the first benchmark for aerospace TSAD. \benchmark comprises nine tasks that combine three pattern-wise anomaly types, univariate and multivariate signals, and both in-loop and out-of-loop feedback scenarios, yielding 108,000 data points. Using this benchmark, we systematically evaluate state-of-the-art open-source LLMs under two paradigms: \Directly, which labels anomalies within sliding windows, and \Predictionbased, which detects anomalies from prediction errors. To reflect operational needs, we reformulate evaluation at the window level and propose three user-oriented metrics: Alarm Accuracy (AA), Alarm Latency (AL), and Alarm Contiguity (AC), which quantify alarm correctness, timeliness, and credibility. We further examine two enhancement strategies, few-shot learning and retrieval-augmented generation (RAG), to inject domain knowledge. The evaluation results show that (1) LLMs perform well on univariate tasks but struggle with multivariate telemetry, (2) their AA and AC on multivariate tasks approach random guessing, (3) few-shot learning provides modest gains whereas RAG offers no significant improvement, and (4) in practice LLMs can detect true anomaly onsets yet sometimes raise false alarms, which few-shot prompting mitigates but RAG exacerbates. These findings offer guidance for future LLM-based TSAD in aerospace software.

\end{abstract}


\section{Introduction}
\label{sec:Introduction}
With the rapid growth and increasing complexity of modern software systems, ensuring their safety and reliability has become a critical concern. Among various assurance techniques, time series anomaly detection (TSAD) plays a vital role in monitoring software system health by identifying abnormal behaviors in operational data. Effective anomaly detection enables timely responses to system faults, performance degradation, or potential security attacks. Conversely, failure to detect anomalies promptly can result in cascading failures, prolonged downtime, and severe economic losses as well as safety risks, particularly in large-scale, safety-critical scenarios.

\begin{figure}[t]
    \centering
    \includegraphics[width=0.5\textwidth]{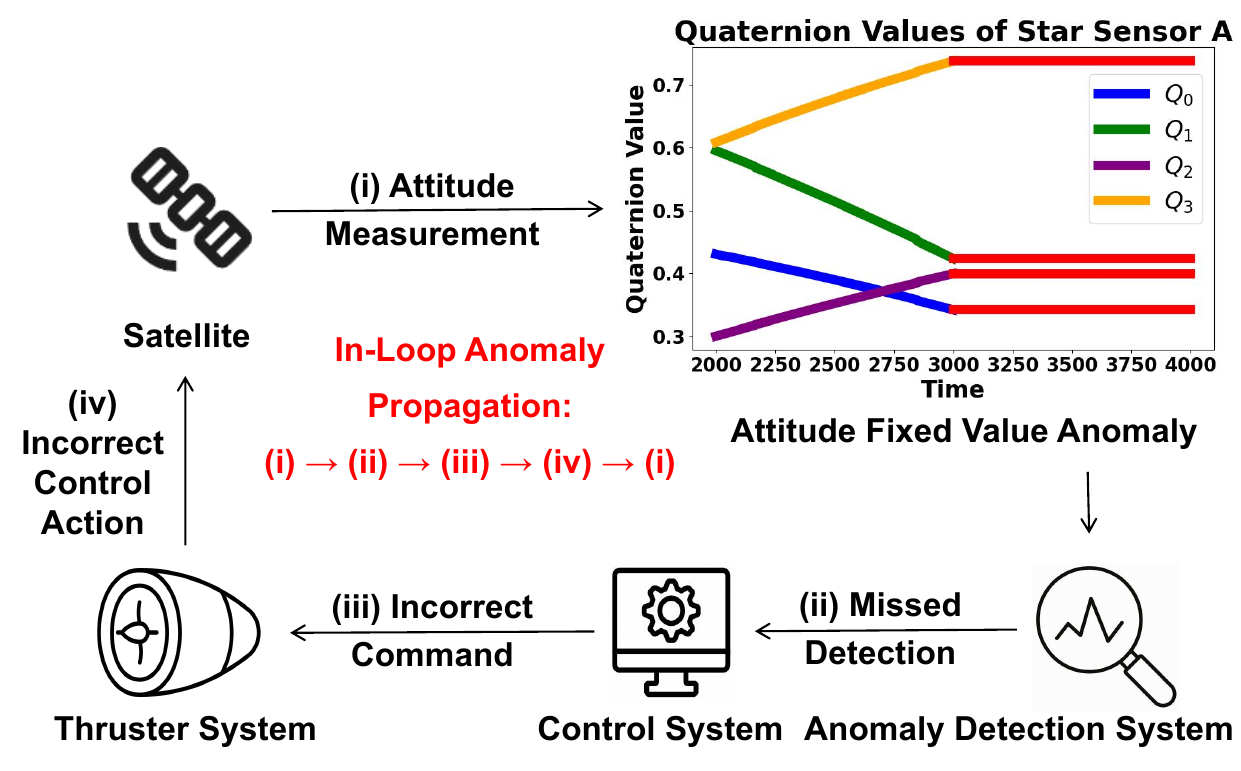}
    \caption{A missed detection of a fixed-value anomaly in star sensor attitude measurements can propagate through the control loop, causing erroneous commands and continuous thruster firing, ultimately resulting in critical satellite damage.}
    \label{fig:anomaly_failure}
\end{figure}
Deep learning has become the dominant approach for TSAD. Supervised approaches, which require labeled data, have achieved significant success but are limited by the difficulty of obtaining sufficient labeled anomalous data in real-world systems\cite{DBLP:conf/aaai/JangK25,DBLP:conf/icde/Liu0C0WWZ23,DBLP:conf/cloud/ZimmermanKSCFBK19,DBLP:conf/kdd/PangSH19,DBLP:journals/symmetry/QiuDQ19,DBLP:journals/kbs/Villa-PerezCLMV21}. This limitation has led to the rise of unsupervised approaches, which detect anomalies by learning normal patterns from unlabeled data\cite{chen2022adaptive,he2022share,zhang2023fkpiselect,DBLP:conf/issre/YuPZWLXP23,jing2024diner,liu2024uac,tang2024mlp,wang2024revisiting,lee2023maat,liu2023practical,kim2024model,liu2025gcad,huang2025graph,chen2022deep,DBLP:conf/ijcai/ZhangZT22,yue2024sub,DBLP:conf/ijcai/LiuHZLM24,DBLP:conf/nips/WangZQ0W0L23,DBLP:conf/nips/SongKOC23,DBLP:conf/nips/LaiSGLB23,DBLP:journals/tkde/LiYWJY23,DBLP:journals/tkde/LiuZDHZSC23,DBLP:journals/tkde/ZhangCWP23,DBLP:journals/tkde/ZhengMZZZYYZJ24,DBLP:journals/tkde/XuWJLWP24,DBLP:journals/tkde/KimKML24,DBLP:journals/tkde/ZhangXCCZXY24}. However, these unsupervised approaches often struggle with generalization and transferability across diverse time series, requiring separate models to be trained for each individual series. This results in unaffordable maintenance costs in large-scale systems where numerous time series need to be monitored\cite{si2024timeseriesbench}.

Recent advances have demonstrated the application of Large Language Models (LLMs) to time series anomaly detection, despite these models originally being designed for natural language processing tasks~\cite{DBLP:conf/nips/GaoKQHTZ24,zhou2023one,DBLP:conf/ijcai/LiuHZLM24}. These models exhibit strong generalization capabilities, eliminating the need for training on specific time series and offering a promising, training-free alternative approach. Alnegheimish et al.~\cite{alnegheimish2024large} introduce the SIGLLM framework, which adapts LLMs through two paradigms: \Directly and \Predictionbased, showing that zero-shot LLMs can successfully find anomalies in univariate time series. To improve the anomaly detection performance and interpretability of LLMs, \cite{DBLP:journals/corr/abs-2405-15370} proposed the LLMAD framework for univariate time series. This framework enhances LLMs by leveraging domain knowledge injection and Chain-of-Thought (CoT). To the best of our knowledge, no work has explored the anomaly detection capabilities of LLMs on multivariate time series to date.

Despite their potential, applying LLMs to TSAD in aerospace is largely unexplored, facing three fundamental challenges:
\textbf{(1) Complexity of Aerospace Data:} Aerospace telemetry exhibits highly diverse anomaly types and complex inter-variable relationships, governed by physical laws and feedback loops. Anomalies may propagate through the system (e.g., an undetected fixed-value anomaly in a star sensor leading to incorrect control actions and potential satellite damage, as illustrated in Fig.~\ref{fig:anomaly_failure}). Such complexity makes effective anomaly detection particularly challenging for generic LLMs.
\textbf{(2) Misalignment of Evaluation Metrics:} Widely used point-based classification metrics, such as F1-score, do not reflect operational practice. In real scenarios, LLMs analyze windows of telemetry data, and users are concerned with whether a window contains actionable anomalies, rather than individual point-wise labels. Thus, current metrics do not adequately measure practical detection performance.
\textbf{(3) Lack of Domain Knowledge:} LLMs, trained primarily on general-purpose data, lack the domain-specific understanding necessary for aerospace applications, such as physical control laws, sensor failure modes, and operational procedures, which limits their ability to detect and interpret critical anomalies.
To assess the performance of LLMs on diverse anomaly types in aerospace, we construct \benchmark, the first comprehensive benchmark specifically designed for aerospace TSAD. \benchmark encompasses nine representative tasks covering fixed-value, constant deviation, and time-varying deviation anomalies, across both univariate (U) and multivariate (M) signals, and includes both in-loop (M-IL) and out-of-loop (M-OL) feedback scenarios, totaling 108,000 data points. Using this benchmark, we systematically evaluate state-of-the-art open-source LLMs under two principal paradigms: \Directly, which involves identifying anomalies within windows, and \Predictionbased, which detects anomalies based on prediction errors.
To more closely align evaluation with practical user needs, we recast the assessment as a window-based task and introduce three novel user-oriented metrics: Alarm Accuracy (AA), Alarm Latency (AL), and Alarm Contiguity (AC). These metrics collectively capture the correctness, timeliness, and reliability of alarms from an operational standpoint. Additionally, we explore enhancement strategies for LLMs, including few-shot learning and retrieval-augmented generation (RAG), evaluating their impact on anomaly detection performance with conventional and proposed metrics.

Our evaluation yields four main findings: 
First, LLMs perform notably well in univariate
tasks, though they still lag behind SOTA unsupervised
approaches, and they show a slight advantage in M-IL over M-OL; the \Predictionbased paradigm consistently outperforms the \Directly paradigm.
Second, standard point-wise metrics fail to reflect operational requirements. Our proposed window-based metrics (AA, AL, AC) enable a more realistic and actionable assessment of anomaly detection for aerospace applications.
Third, Few-shot learning can enhance LLM performance under the \Directly paradigm, while RAG provides only marginal or inconsistent improvements. 
Finally, a case study on a real in-orbit anomaly was conducted, showing that LLMs are capable of providing timely anomaly alarms, but may generate false positives, and few-shot learning can help mitigate false alarms.

The contributions of this paper are as follows:
\begin{itemize}
    \item We present the first systematic evaluation of large language models for time series anomaly detection in aerospace applications. To facilitate this, we introduce \benchmark, a comprehensive benchmark comprising nine tasks that encompass three pattern-wise anomaly types and three control-loop scenarios, enabling rigorous comparison with SOTA unsupervised approaches.
    
    \item We propose three user-oriented window-level metrics (AA, AL, AC) for practical evaluation to assess the correctness, timeliness, and credibility of alarms from an operational perspective.
    
    \item We investigate enhancement strategies, including few-shot learning and RAG, and validate their effectiveness through experiments and a real-world anomaly case study, revealing their practical benefits and limitations for LLM-based detection in aerospace contexts.
    
\end{itemize}                                                                 

\begin{figure*}[!tb]
  \centering
  \includegraphics[width=\linewidth]{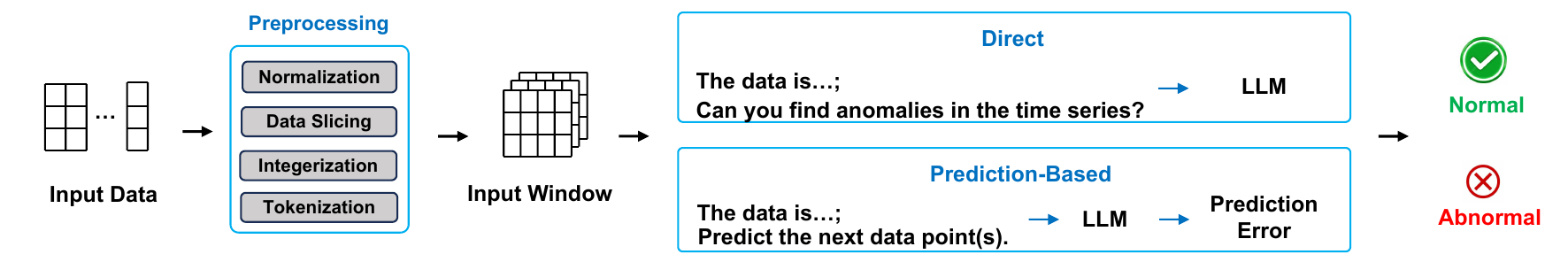}
  \caption{Pipeline of time series anomaly detection with zero-shot LLMs}
  \label{fig:llm-based}
\end{figure*}

\section{Background and Related Work}
\subsection{Time Series Anomaly Detection}
Time series anomaly detection refers to the task of identifying data points or subsequences within a temporal sequence that deviate significantly from the expected normal behavior. In this context, an anomaly is a data point or interval whose characteristics differ markedly from the majority of the time series. 
This task is essential in software reliability engineering and has been dominated by deep learning approaches, which leverage sophisticated models to capture temporal dependencies and inter-variable correlations. Approaches can generally be categorized as supervised or unsupervised, depending on the availability of labeled data.
\subsubsection{Supervised Anomaly Detection}
Supervised approaches cast anomaly detection as a classification problem, requiring a labeled training dataset $\mathcal{T}$ where each data point $\mathbf{x}_t\in \mathcal{T}$ is annotated as normal or anomalous. The model is trained to distinguish between these classes and subsequently labels each test data point during inference. While effective, this approach is often impractical for real-world software systems due to the scarcity of anomalies and the high cost of manual labeling.

\subsubsection{Unsupervised Anomaly Detection}
\label{sec:Unsupervised Anomaly Detection}
Unsupervised approaches do not rely on labeled data. These approaches typically assume anomalies are rare within the training set $\mathcal{T} = \{\mathbf{x}_1, \mathbf{x}_2, \ldots, \mathbf{x}_N\}$ and treat most data as normal. Both training and test data are segmented into overlapping windows via a sliding window strategy. Models are trained to learn normal behavior and, during inference, assign an anomaly score to each data point, reflecting its deviation from expected patterns. Data points with scores exceeding a predefined threshold are flagged as anomalous. Unsupervised anomaly detection approaches can be further categorized according to how anomaly scores are generated:

\begin{packeditemize}
\item  \textbf{\Directly paradigm:} These approaches detect anomalies by directly assigning an anomaly score to each data point within a window, based on its conformity to learned normal patterns. Representative approaches include:
\begin{packeditemize}
    \item \textbf{Reconstruction-based approach:} It encodes input windows into low-dimensional representations and reconstructs them. High reconstruction errors indicate anomalies \cite{li2023tadl,jing2024diner,liu2024uac,kim2024model,wang2024revisiting,huang2025graph,yue2024sub}.
    \item \textbf{Contrastive learning-based approach:} It learns representations where normal and anomalous data points are well-separated by using contrastive objectives during training. Samples that deviate from learned normal patterns receive higher anomaly scores \cite{li2023deepdiscord,fang2024temporal}.
\end{packeditemize}
\item \textbf{\Predictionbased paradigm:} These approaches forecast future data points based on the current window and classify points as anomalies if the prediction error is significant \cite{lee2023maat,liu2023practical,zhang2024dyntrackr,liu2025gcad,si2023beyond}.
\end{packeditemize}

\subsubsection{Zero-Shot Anomaly Detection with LLMs}
\label{sec:Zero-Shot Anomaly Detection with LLMs}
LLMs with extensive pre-training and remarkable generalization capabilities have recently enabled innovative approaches to time series anomaly detection. Leveraging zero-shot learning, LLMs can identify anomalies in sequential data without requiring task-specific training or labeled examples. As shown in Fig.~\ref{fig:llm-based}, the zero-shot workflow comprises two primary steps: preprocessing and anomaly detection:
\begin{packeditemize}
    \item \textbf{Preprocessing} includes normalization, slicing the time series into windows, integerization (rescaling to integer values), and tokenization. For LLMs such as GPT~\cite{DBLP:journals/corr/abs-2305-00118} and DeepSeek~\cite{DBLP:journals/corr/abs-2412-19437}, digits are separated by spaces for accurate tokenization~\cite{gruver2023large, DBLP:journals/corr/abs-2305-14201}; in contrast, LLaMA~\cite{DBLP:journals/corr/abs-2307-09288} and Qwen~\cite{DBLP:journals/corr/abs-2505-09388} support direct numerical tokenization.
    \item \textbf{Anomaly Detection} is performed via two paradigms as follows~\cite{alnegheimish2024large}:
    \begin{packeditemize}
        \item \textbf{\Directly paradigm:} The LLM is prompted to identify anomalies directly within each window, returning indices of anomalous data points.
        \item \textbf{\Predictionbased paradigm:} The LLM predicts the next data point(s) according to the input window, and deviations between predictions and actual values are used to detect anomalies.
    \end{packeditemize}
\end{packeditemize}

\subsection{Related Work}
Recent years have seen extensive empirical research on time series anomaly detection. Audibert et al.\cite{audibert2022deep} provided a comparative study encompassing traditional, machine learning, and deep neural network approaches, while Li et al.\cite{li2023empirical} conducted a comprehensive evaluation of unsupervised deep learning algorithms. Sarfraz et al.\cite{sarfraz2024position} critically examined state-of-the-art deep learning techniques, benchmarking them against simpler baselines. Additionally, Freeman et al.\cite{DBLP:conf/ijcai/FreemanMB022} offered a broad survey of twelve algorithms, ranging from statistical models to early neural network approaches, assessed across diverse time series characteristics.

With the advent of LLMs, research attention has shifted toward their potential in anomaly detection~\cite{DBLP:conf/ijcai/LiuHZLM24,DBLP:conf/nips/GaoKQHTZ24,zhou2023one}. Recent work~\cite{alnegheimish2024large, DBLP:journals/corr/abs-2405-15370} have demonstrated that LLMs can detect anomalies with minimal training, but primarily evaluated univariate time series. In contrast,our work is the first to systematically investigate LLMs for multivariate time series anomaly detection, particularly in complex industrial settings such as multi-sensor aerospace telemetry. This fills a critical gap by assessing whether LLMs can model the intricate inter-variable dependencies inherent in real-world systems.

Beyond detection algorithms, evaluation methodology has received increasing attention. Si et al.\cite{si2024timeseriesbench} proposed the “All-in-One” paradigm for unified model assessment, while Puccetti et al.\cite{puccetti2024detection} stressed the importance of detection latency. Kim et al.~\cite{kim2022towards} cautioned that the widely used “point adjustment” protocol can artificially inflate performance scores. In response, we introduce novel, user-oriented evaluation metrics specifically designed for aerospace operational contexts, capturing not only detection accuracy but also the timeliness and credibility of anomaly alerts, which are key considerations in safety-critical applications.

Finally, the choice of evaluation datasets is essential for rigorous benchmarking. Wu et al.\cite{wu2021current} identified limitations in many mainstream datasets and released the UCR Time Series Anomaly Archive as a more robust alternative, while Si et al.\cite{si2024timeseriesbench} provided the industrial-scale TimeSeriesBench dataset to reflect real-world complexity. In contrast, we construct a new benchmark dataset derived from authentic aerospace telemetry, encompassing diverse anomaly types and both out-of-loop and in-loop scenarios. This application-driven design ensures that our evaluation aligns with the actual challenges faced in industrial practice, setting our work apart from previous studies on generic datasets.

\section{\benchmark Construction}
\subsection{Types of Time Series Anomalies in Aerospace}
\label{sec:Taxonomy of Time Series Anomalies in Aerospace}

\begin{figure}[!tb]
    \centering
    \begin{subfigure}[b]{0.24\textwidth}
        \centering
        \includegraphics[width=\textwidth]{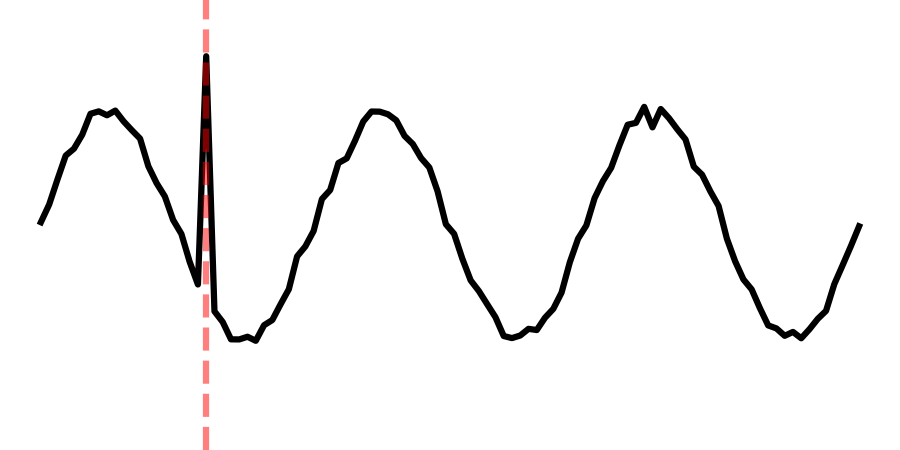}
        \caption{Point-Wise Anomaly}
        \label{fig:point_wise_anomaly}
    \end{subfigure}
    \hfill
    \begin{subfigure}[b]{0.24\textwidth}
        \centering
        \includegraphics[width=\textwidth]{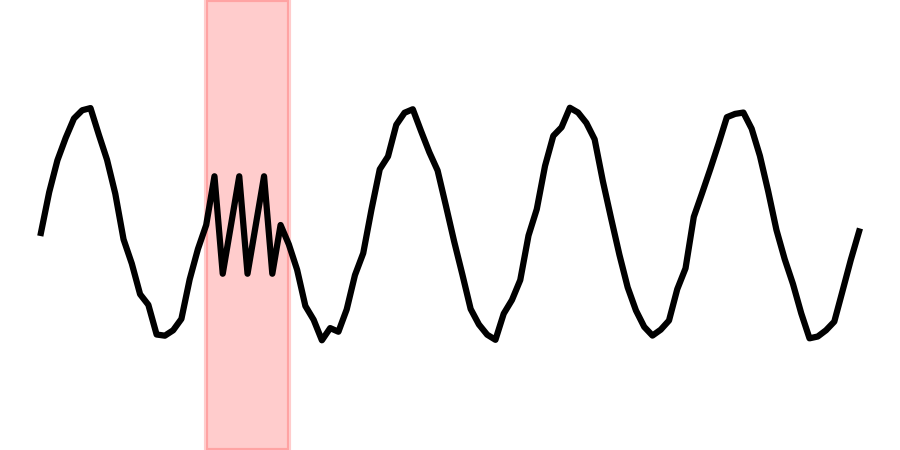}
        \caption{Pattern-Wise Anomaly}
        \label{fig:pattern_wise_anomaly}
    \end{subfigure}
    \caption{Overview of time series anomaly types.}
    \label{fig:two_anomaly}
\end{figure}

Following the behavior-driven taxonomy of \cite{lai2021revisiting}, time series anomalies in aerospace systems are broadly classified as \textbf{point-wise} or \textbf{pattern-wise} (see Fig.\ref{fig:two_anomaly}). Point-wise anomalies are isolated events, such as spikes or glitches, occurring at single time steps. In contrast, pattern-wise anomalies span contiguous intervals, characterized by sustained deviations from normal trends \cite{si2024timeseriesbench}. While point-wise anomalies are typically addressed using conventional out-of-limits (OOL) checks or basic statistical approaches\cite{5633180,https://doi.org/10.1609/aimag.v35i4.2553,1659593}, pattern-wise anomalies pose greater detection challenges in aerospace applications. This study thus focuses on the evaluation of LLM-based approaches for complex pattern-wise anomalies.

\begin{figure}[!tb]
    \centering
    \begin{subfigure}[t]{0.15\textwidth}
        \centering
        \includegraphics[width=\textwidth]{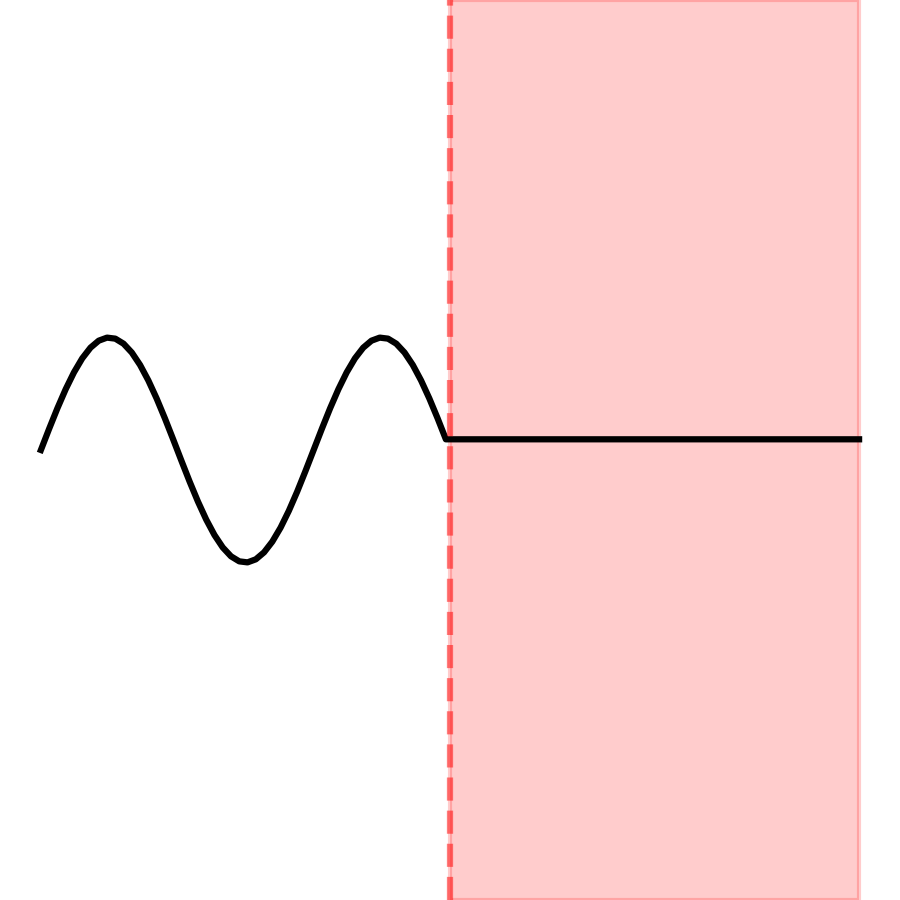}
        \caption{FVA}
        \label{fig:constant_anomaly}
    \end{subfigure}
    \hfill
    \begin{subfigure}[t]{0.15\textwidth}
        \centering
        \includegraphics[width=\textwidth]{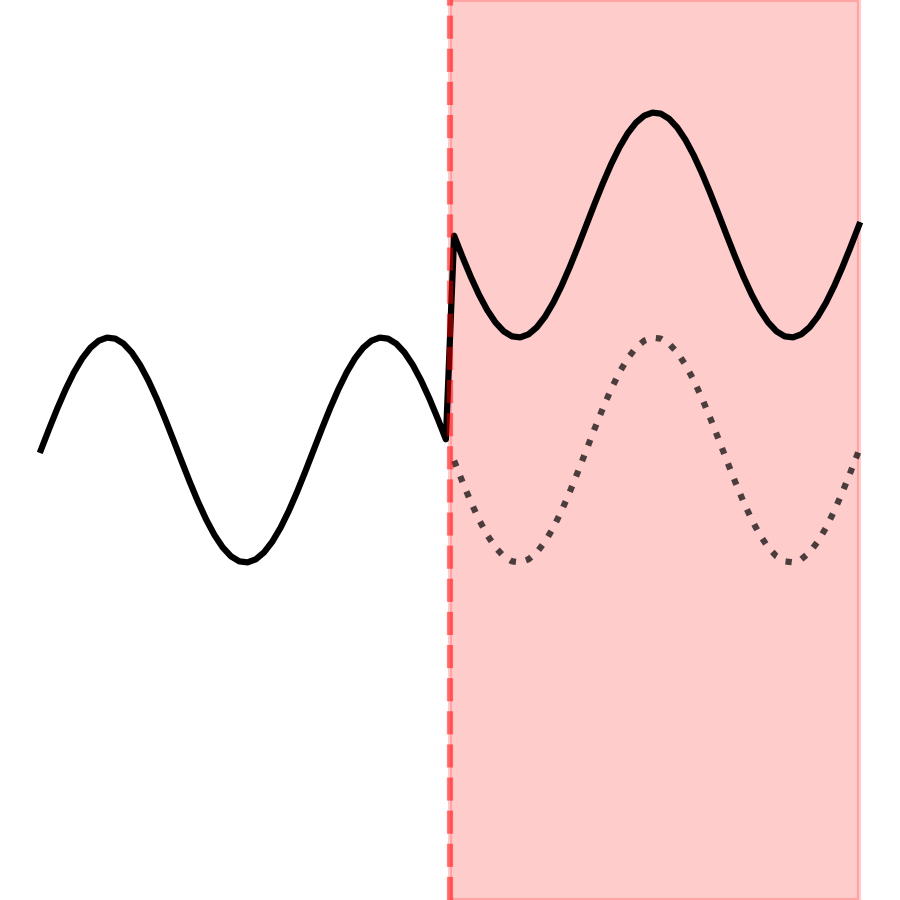}
        \caption{CDA}
        \label{fig:constant_deviation_anomaly}
    \end{subfigure}
    \hfill
    \begin{subfigure}[t]{0.15\textwidth}
        \centering
        \includegraphics[width=\textwidth]{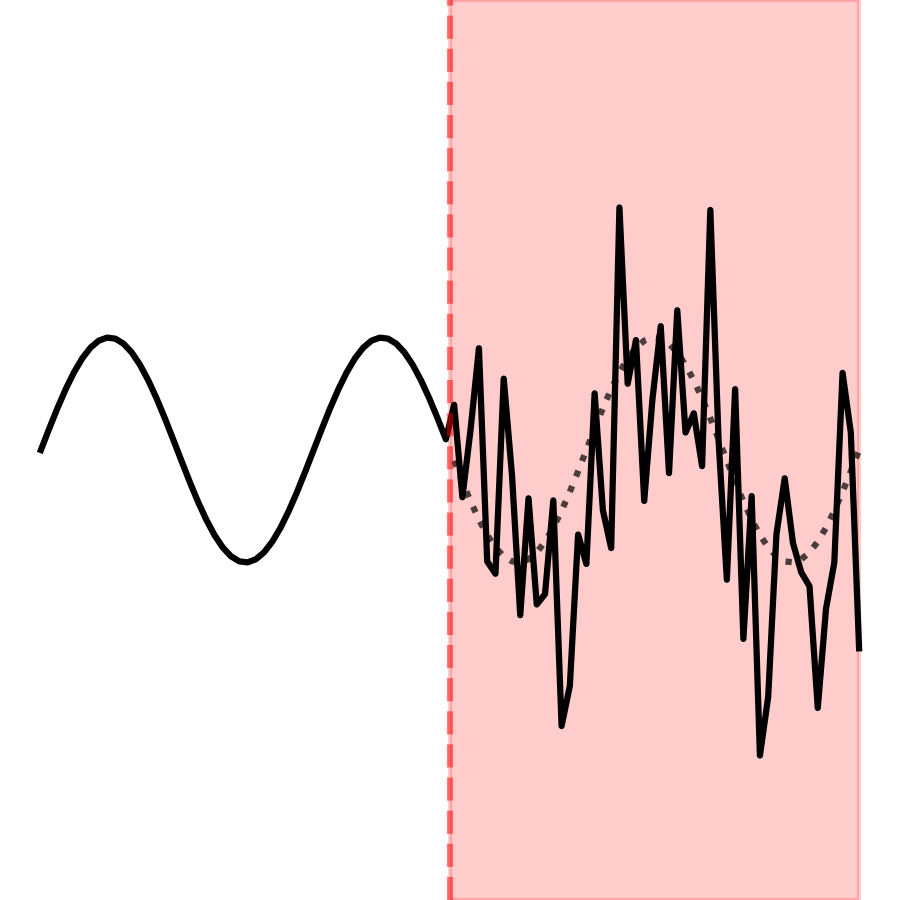}
        \caption{TVDA}
        \label{fig:time_varying_deviation_anomaly}
    \end{subfigure}
    \caption{Visualization of three types of pattern-wise anomaly.}
    \label{fig:Anomaly Classification}
\end{figure}

Within the pattern-wise category, we propose an aerospace-specific taxonomy consisting of two principal anomaly types, defined as follows (illustrated in Fig.\ref{fig:Anomaly Classification}). Let $x(t)$ denote the expected output at time $t$ and $y(t)$ the observed anomalous output for $t\in[t_1, t_2]$:
\begin{packeditemize}
    \item \textbf{Fixed-Value Anomaly (FVA)}: 
    The output is frozen at its initial anomalous value: $y(t) = x(t_1)$, $\forall t \in [t_1, t_2]$ (Fig.~\ref{fig:constant_anomaly}).
    \item \textbf{Deviation Anomaly}: 
     The output is offset from the normal trajectory: $y(t) = x(t) + \Delta$. This is subdivided as:
    \begin{packeditemize}
        \item \textbf{Constant Deviation Anomaly (CDA)}: 
        $\Delta = b$, a constant offset (Fig.\ref{fig:constant_deviation_anomaly}).
        \item \textbf{Time-Varying Deviation Anomaly (TVDA)}: \
        $\Delta = \delta(t)$, a dynamic, time-dependent deviation (Fig.~\ref{fig:time_varying_deviation_anomaly}).
    \end{packeditemize}
\end{packeditemize}


\subsection{Evaluation Task Definition}
To comprehensively assess anomaly detection approaches in aerospace scenarios, our benchmark comprises nine tasks along two axes: signal dimensionality and anomaly propagation. The tasks are defined as follows:
\begin{packeditemize}

    \item \textbf{Univariate (U) Tasks:}
    Evaluates detection based solely on temporal patterns in a single variable.
    \item \textbf{Multivariate (M) Tasks:}
    Assesses reasoning over dependencies among multiple variables, which are intrinsic to aerospace systems. To capture the diverse propagation effects of anomalies within such interconnected environments, this category is further subdivided based on the anomaly’s influence on system-wide behavior:
    \begin{packeditemize}
    \item \textbf{In-Loop (IL) Tasks:} 
    Anomalies affect sensors engaged in closed-loop control, with fault effects propagating system-wide (see Fig.~\ref{fig:anomaly_failure}).
    \item \textbf{Out-of-Loop (OL) Tasks:} 
    Anomalies occur in backup or inactive sensors, remaining isolated without affecting system-level behavior.
    \end{packeditemize}
\end{packeditemize}

Each of these three settings, i.e., Univariate, Multivariate In-Loop, and Multivariate Out-of-Loop, is combined with the three previously defined pattern-wise anomaly types (FVA, CDA, and TVDA), resulting in nine evaluation tasks.

\subsection{\benchmark Construction}
We present \benchmark, a comprehensive benchmark for time series anomaly detection in aerospace software. The construction pipeline comprises scenario design (including anomaly injection), data processing, and dataset formulation.

\begin{table}[!tb]
\centering
\footnotesize
\caption{The statistics of the constructed \benchmark}
\begin{tabular}{c|c|c|c}
\toprule
\textbf{Signal Dimensions} & \textbf{Tasks} & \textbf{Data Points} & \textbf{Anomaly Ratio} \\ \midrule
\multirow{3}{*}{Univariate} & U-FVA & 8000 &  25\% \\
 & U-CDA & 8000 & 25\% \\
 & U-TVDA & 8000 &  25\% \\ \midrule
\multirow{6}{*}{Multivariate} & M-IL-FVA & 8000 & 25\% \\
 & M-IL-CDA & 8000 & 25\%  \\
 & M-IL-TVDA & 8000 & 25\%  \\
 & M-OL-FVA & 8000 & 25\%  \\
 & M-OL-CDA & 8000 & 25\%  \\
 & M-OL-TVDA & 8000 & 25\%  \\ 
 \midrule
\multirow{2}{*}{Total} & Training Set & 72000 & 0\%  \\
 & Test Set & 36000 & 50\% \\ 
 \bottomrule
\end{tabular}
\label{tab:task_stat}
\end{table}

\subsubsection{Scenario Design and Anomaly Injection.} 
The nine evaluation tasks span both univariate and multivariate domains, with the latter further divided into in-loop and out-of-loop contexts depending on sensor involvement in feedback control.
For multivariate in-loop tasks, anomalies are injected into active sensors (i.e., Star Sensor 1, Gyroscope 1) whose outputs impact control actions. In out-of-loop tasks, faults affect redundant sensors with no influence on control, ensuring anomaly isolation. Each task is instantiated with two simulation runs (one each for star sensor and gyroscope faults), yielding 12 experiments in total. Each experiment lasts 2000 seconds: 1500 seconds normal operation followed by 500 seconds with injected faults. Anomalies are introduced as follows:
\begin{packeditemize}
    \item \textbf{FVA}: sensor output is frozen at a constant value.
    \item  \textbf{CDA}: a fixed offset is added to the true signal.
    \item  \textbf{TVDA}: the output gradually diverges using a time-dependent transformation (e.g., misalignment matrix for star sensors or drifting noise for gyroscopes).
\end{packeditemize}

All injected anomalies were validated by domain experts and reproduce realistic on-orbit fault behaviors.

\subsubsection{Dataset Preprocessing}
All raw telemetry data were recorded at 2 Hz (one data point every 0.5 seconds). To ensure high data quality and validity of the benchmark, two primary preprocessing steps were applied:
(i) Correction of sign inversions in quaternion-based star sensor outputs, addressing inherent sign ambiguity that can cause artificial discontinuities and be misinterpreted as anomalies;
(ii) Removal of the initial stabilization phase following simulation startup, thereby excluding transient behaviors unrelated to fault dynamics and focusing the analysis on steady-state operations.

\subsubsection{\benchmark Formulation} 
Table~\ref{tab:task_stat} summarizes the overall statistics of the constructed \benchmark, which covers both univariate and multivariate anomaly detection tasks, each further subdivided by anomaly type and system context. Each data point in the dataset contains:
\begin{packeditemize}
\item \textbf{Sensor Telemetry Data}: Measurements or parameters collected from system sensors, sampled at 2 Hz.
\item \textbf{Task Label}: The specific anomaly detection task, indicating both the anomaly pattern and the system configuration.
\item \textbf{Ground Truth Annotation}: A binary label indicating whether the data point is normal or anomalous.
\end{packeditemize}


For each task, the dataset contains a fixed number of data points (8,000 per task), as detailed in Table~\ref{tab:task_stat}. This uniformity across tasks is a result of the controlled simulation settings, which generate consistent sequence lengths and facilitate systematic comparison between different scenarios. Such standardization ensures that evaluation results are directly comparable and not confounded by differences in data volume or segment length.
The benchmark is divided into a training set and a test set. The training set includes only normal data, with no injected anomalies, providing a clean baseline for model training. In contrast, the test set comprises both normal and anomalous segments, with an overall anomaly ratio of 50\%, enabling rigorous and reproducible assessment of detection performance under challenging conditions. In total, the benchmark comprises 108,000 data points.

\section{Study Design}


To systematically evaluate LLMs for time series anomaly detection in aerospace applications, we designed a study guided by four research questions, with a focus on industrial relevance and practical deployment.
\subsection{Research Questions}

\textbf{RQ1: How do LLMs perform in anomaly detection on \benchmark compared to SOTA unsupervised approaches?}
We evaluate LLMs across diverse tasks using standard metrics (F1, AUROC, AUPRC) to benchmark their anomaly detection capabilities in aerospace contexts.

\textbf{RQ2: Can LLMs meet practical operational requirements for alarm accuracy, timeliness, and credibility in mission-critical aerospace scenarios?}
We assess whether LLMs can generate alarms that are both timely and trustworthy, as measured by user-oriented, window-level metrics essential for real-world deployment.


\textbf{RQ3: How much can LLM performance be improved through enhancement strategies such as few-shot learning and RAG?}
We examine the value of integrating domain knowledge via few-shot prompting and reference retrieval for boosting anomaly detection effectiveness, as these simple and widely used methods provide a clear starting point for exploring enhancement strategies.

\textbf{RQ4: How do LLMs perform on real-world aerospace anomaly detection tasks?}
We validate the practical applicability of LLM-based detection through a case study on an actual in-orbit anomaly event.

\subsection{Evaluation Metrics}
\label{sec:Evaluation Metrics}
\textbf{Standard Classification Metrics.}
To benchmark the anomaly detection capabilities of LLMs and unsupervised approaches, we adopt the F1-score, a widely used metric for time series anomaly detection~\cite{DBLP:conf/ecai/LiZCZY23, DBLP:conf/nips/WangZQ0W0L23, DBLP:conf/issre/SunGLWKZLXP24}. Anomalies are detected by treating each data point as either \textit{normal} or \textit{abnormal}, allowing direct use of binary classification metrics. Additionally, to mitigate the impact of threshold selection inherent in \Predictionbased approaches, we further use two threshold-independent metrics: the Area Under the Receiver Operating Characteristic Curve (AUROC) and the Area Under the Precision-Recall Curve (AUPRC), both calculated on the continuous prediction error~\cite{10.1609/aaai.v38i12.29210, huang2025graph}.

\textbf{User-oriented Metrics.} 
As discussed  in~\cref{sec:Introduction}, standard point-level metrics often fail to capture the operational needs of aerospace systems users. To address this, we aggregate model outputs at the window level. For each input window, the LLM produces binary labels for all data points; we count the number of points flagged as anomalous. If this count exceeds a set threshold, the window is labeled as anomalous, which directly reflects the user’s concern: determining if a window of time contains an anomaly requiring attention. We then introduce three metrics that align with industrial requirements:

\begin{figure}[!tb]
    \centering
    \begin{subfigure}[t]{0.45\textwidth}
        \centering
        \includegraphics[width=\textwidth]{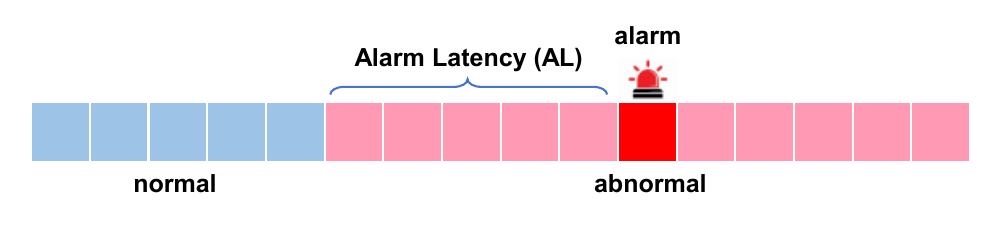}
        \caption{AL measures the delay in terms of the number of input windows from anomaly onset to the initial alarm}
        \label{fig:alarm_latency}
    \end{subfigure}
    \vfill
    \begin{subfigure}[t]{0.45\textwidth}
        \centering
        \includegraphics[width=\textwidth]{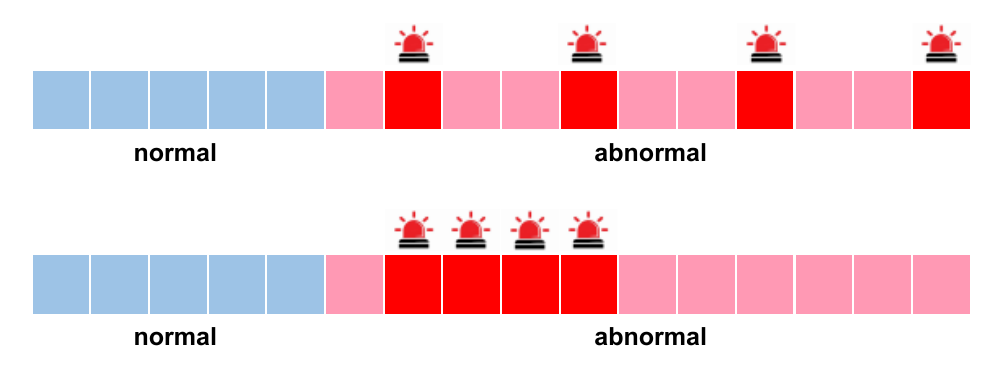}
        \caption{AC measures the degree of alarm concentration, distinguishing between dispersed (top) and contiguous (bottom) alarms.}
        \label{fig:alarm_contiguity}
    \end{subfigure}
    \hfill
    \caption{Conceptual illustration of the AL and AC metrics.}
    \label{fig:Practical Metrics}
\end{figure}

\textit{Alarm Accuracy (AA).} The primary goal of this metric is to measure the correctness of window-level alarm decisions:
\[
AA = \frac{TP + TN}{TP + TN + FP + FN}
\]
where TP, TN, FP, FN are the counts of true/false positives/negatives at the window level.

\textit{Alarm Latency (AL).} We propose the AL metric to quantify the average delay, in number of windows, from the true onset of an anomaly to the first triggered alarm. As shown in Fig.~\ref{fig:alarm_latency}, each input window is visualized as a colored block: blue indicates a normal input window, pink denotes an abnormal window where the detector fails to raise an alarm, and red signifies an abnormal window where the detector successfully raises an alarm.
AL quantifies the delay between the start of an anomalous event and the first window in which an alarm is successfully triggered. It is measured by the number of windows. Lower AL indicates more timely detection. 

%
\textit{Alarm Contiguity (AC).} As illustrated in Fig.~\ref{fig:alarm_contiguity}, the AC metric quantifies how concentrated or continuous alarms are within an anomalous period, directly reflecting the trustworthiness and usability of the alarm system for users. In aerospace scenarios, highly contiguous alarms, those sustained throughout an anomaly, are much more actionable and reassuring than sporadic or scattered alarms. Contiguous alarms reliably signal ongoing risk, minimizing the chance that users will miss critical events, while dispersed alarms may cause confusion or erode user trust.

For each anomalous segment, defined by consecutive input windows $[s_n, e_n]$ for $n \in [1, N]$ (where $N$ is the total number of anomalous segments), we identify $L_n$ contiguous alarm runs, with $[\alpha_i, \beta_i]$ denoting the $i$-th alarm run and $i \in [1, L_n]$. The length of the longest alarm run is normalized by the segment length. The overall AC score is then given by:
\[
AC = \frac{1}{N} \sum_{n=1}^{N} \frac{\max_{i\in[1,L_n]} (\beta_i - \alpha_i)}{e_n - s_n}
\]
A higher AC indicates more sustained and reliable alarms, which are crucial for user trust and timely response in aerospace applications.

\subsection{Evaluated Approaches}
\textbf{Selected Unsupervised learning Approaches.}
Guided by recent empirical studies and surveys~\cite{si2024timeseriesbench,puccetti2024detection,sarfraz2024position,li2023empirical,kim2022towards,audibert2022deep,DBLP:conf/ijcai/FreemanMB022,wu2021current,jin2024survey,liu2025gcad,yue2024sub,fang2024temporal}, we selected state-of-the-art unsupervised anomaly detection approaches representative of two main paradigms as mentioned in~\cref{sec:Unsupervised Anomaly Detection}: Sub-Adjacent~\cite{yue2024sub} and TFMAE~\cite{fang2024temporal} for the \Directly paradigm, and GCAD~\cite{liu2025gcad} for the \Predictionbased paradigm.

\textbf{Studied LLMs.}
For anomaly detection with large language models, we evaluated two open-source SOTA LLMs: DeepSeek-V3-0324~\cite{DBLP:journals/corr/abs-2412-19437} (DeepSeek-V3) and Qwen3-235B-A22B~\cite{DBLP:journals/corr/abs-2505-09388} (Qwen3). Both models were obtained from the Hugging Face model hub\footnote{\url{https://huggingface.co/models}} and run locally via API. We did not include commercial or closed-source models (such as the GPT series) in our evaluation, as they cannot be deployed for local inference. This limitation is particularly important given the data confidentiality requirements in the aerospace domain, which prohibit uploading proprietary telemetry data to third-party cloud services.

\begin{table}[!tb]
\footnotesize
\setlength{\tabcolsep}{3.5pt}
\centering
\caption{Window-related hyperparameters Configurations}
\label{tab:window_configs}
\begin{tabular}{cccccc}
\toprule
\textbf{Paradigm} & \textbf{Approach} & \textbf{\makecell{Signal \\ Dimensions}} & \textbf{\makecell{Window \\ Size}} & \textbf{\makecell{Step \\ Size}} & \textbf{\makecell{Prediction \\ Horizon}} \\
\midrule
\multirow{4}{*}{Directly} & \multirow{2}{*}{\makecell{Zero-shot LLMs}} & U & 500 & 100 & -- \\
\cmidrule{3-6}
~& ~ & M & 10 & 10 & -- \\
\cmidrule{2-6}
 ~& Sub-Adjacent & U\&M & 100 & 1 & -- \\
\cmidrule{2-6}
~& TFMAE & U\&M & 100 & 1 & -- \\
\midrule
\multirow{3}{*}{\makecell{Prediction \\ -based}} & \multirow{2}{*}{\makecell{Zero-shot LLMs}} &U & 1000 & 20 & 20 \\
\cmidrule{3-6}
~&~& M & 20 & 5 & 5 \\
\cmidrule{2-6}
~ & GCAD & U\&M & 30 & 1 & 1 \\
\bottomrule
\end{tabular}
\begin{tablenotes}
\item  $^*$Note that all hyperparameters refer to the number of data points, with an interval of 0.5s between each data point.
\end{tablenotes}
\end{table}

\renewcommand{\arraystretch}{1.5}
\begin{table*}[!tb]
\centering
\caption{RQ1--Performance of zero-shot LLMs and SOTA unsupervised learning approaches on the \benchmark. The best performance within each paradigm (\textit{\Directly} and \textit{\Predictionbased}) is highlighted in \textbf{bold}, and the overall best performance in each column is \underline{\textbf{underlined}}.}
\label{tab:RQ1 performance}
{
\small
\setlength{\tabcolsep}{2pt} 
\begin{tabular}{cccc|c|ccc|c|ccc|c|c}
\cline{1-14}
 & \multicolumn{4}{c}{Univariate} & \multicolumn{4}{c}{Multivariate-Out-of-Loop} & \multicolumn{4}{c}{Multivariate-In-Loop} & Overall \\
\cmidrule(lr){2-5} \cmidrule(lr){6-9} \cmidrule(lr){10-13} \cmidrule(lr){14-14}
Approach & U-FVA & U-CDA & U-TVDA & Avg. & MO-FVA & MO-CDA & MO-TVDA & Avg. & MI-FVA & MI-CDA & MI-TVDA & Avg. & Avg. \\
\hline
\multicolumn{14}{c}{\textit{\Directly}} \\
DeepSeek-V3 & 0.4315 & 0.4278 & \textbf{0.6199} & 0.4931 & 0.5007 & 0.4197 & 0.4355 & 0.4520 & 0.4514 & \textbf{0.4466} & \textbf{0.4912} & 0.4631 & 0.4694 \\
Qwen3 & \underline{\textbf{0.7760}} & 0.3341 & 0.3450 & 0.4850 & 0.2507 & 0.2010 & 0.2847 & 0.2455 & 0.2982 & 0.2583 & 0.2597 & 0.2721 & 0.3342 \\
Sub-Adjacent & 0.5070 & 0.4957 & 0.5050 & 0.5026 & 0.4835 & 0.3905 & 0.4690 & 0.4477 & 0.4770 & 0.4298 & 0.4681 & 0.4583 & 0.4695 \\
TFMAE & 0.5000 & \textbf{0.5850} & 0.5000 & \textbf{0.5283} & \textbf{0.7210} & \textbf{0.8170} & \textbf{0.5145} & \textbf{0.6842} & \textbf{0.7240} & 0.3283 & 0.4760 & \textbf{0.5094} & \textbf{0.5740} \\
\hline
\multicolumn{14}{c}{\textit{\Predictionbased}} \\
DeepSeek-V3 & \textbf{0.5366} & \underline{\textbf{0.7797}} & 0.8192 & \underline{\textbf{0.7118}} & 0.4941 & 0.5846 & 0.5956 & 0.5581 & 0.5041 & 0.6512 & 0.5491 & 0.5681 & 0.6127 \\
Qwen3 & 0.5066 & 0.6517 & \underline{\textbf{0.9722}} & 0.7102 & 0.3931 & 0.4436 & 0.4226 & 0.4198 & 0.4471 & 0.4386 & 0.5246 & 0.4701 & 0.5334 \\
GCAD & 0.0000 & 0.2578 & 0.6981 & 0.3187 & \underline{\textbf{0.9475}} & \underline{\textbf{0.9190}} & \underline{\textbf{0.9560}} & \underline{\textbf{0.9408}} & \underline{\textbf{0.8590}} & \underline{\textbf{0.9300}} & \underline{\textbf{0.9470}} & \underline{\textbf{0.9120}} & \underline{\textbf{0.7238}}\\
\hline
\end{tabular}
}
\end{table*}

\begin{figure*}[!tb]
\begin{subfigure}{0.5\textwidth}
\centering
\begin{tikzpicture}
\def\radius{2cm} 
\def\numaxes{9} 
\foreach \i in {1,...,5} {
    \draw[gray!30, thick] (0,0) circle (\i*\radius/5);
    \pgfmathsetmacro{\labelvalue}{\i*0.2} 
    \node[font=\scriptsize] at (90:\i*\radius/5 + 0.2) {\pgfmathprintnumber[fixed, precision=1]{\labelvalue}};
}
\foreach \angle [count=\i] in {0,40,...,320} {
    \draw[gray!50, thin] (\angle:\radius) -- (0,0);
    \node[font=\scriptsize, align=center] at (\angle:\radius+15) {%
        \ifcase\numexpr\i-1\relax
            U-FVA\or U-CDA\or U-TVDA\or M-OL-FVA\or M-OL-CDA\or M-OL-TVDA\or M-IL-FVA\or M-IL-CDA\or M-IL-TVDA
        \fi};
}
\draw[blue, thick] 
    (0:0.551433*\radius) -- (40:0.721466*\radius) -- (80:0.812333*\radius) --
    (120:0.51116525*\radius) -- (160:0.61556175*\radius) -- (200:0.60026425*\radius) --
    (240:0.534781*\radius) -- (280:0.67156025*\radius) -- (320:0.61766725*\radius) -- cycle;
\draw[red, thick] 
    (0:0.51949825*\radius) -- (40:0.7383145*\radius) -- (80:0.99806725*\radius) --
    (120:0.3579845*\radius) -- (160:0.42155225*\radius) -- (200:0.396617*\radius) --
    (240:0.42860975*\radius) -- (280:0.41528175*\radius) -- (320:0.53722*\radius) -- cycle;
\draw[orange, thick] 
    (0:0.5*\radius) -- (40:0.574*\radius) -- (80:0.768126375*\radius) --
    (120:0.990224*\radius) -- (160:0.97678525*\radius) -- (200:0.993597*\radius) --
    (240:0.91252325*\radius) -- (280:0.98194325*\radius) -- (320:0.98591*\radius) -- cycle;
\begin{scope}[shift={(0,-3cm)}]
    \draw[blue, thick] (-3,0) -- (-2.5,0) node[right, font=\scriptsize] {DeepSeek-V3};
    \draw[red, thick] (-0.5,0) -- (0,0) node[right, font=\scriptsize] {Qwen3};
    \draw[orange, thick] (1.5,0) -- (2,0) node[right, font=\scriptsize] {GCAD};
\end{scope}
\end{tikzpicture}
\caption{AUROC}
\label{fig:radar1}
\end{subfigure}
\hfill
\begin{subfigure}{0.5\textwidth}
\centering
\begin{tikzpicture}
\def\radius{2cm} 
\def\numaxes{9} 
\foreach \i in {1,...,5} {
    \draw[gray!30, thick] (0,0) circle (\i*\radius/5);
    \pgfmathsetmacro{\labelvalue}{\i*0.2} 
    \node[font=\scriptsize] at (90:\i*\radius/5 + 0.2) {\pgfmathprintnumber[fixed, precision=1]{\labelvalue}};
}
\foreach \angle [count=\i] in {0,40,...,320} {
    \draw[gray!50, thin] (\angle:\radius) -- (0,0);
    \node[font=\scriptsize, align=center] at (\angle:\radius+15) {%
        \ifcase\numexpr\i-1\relax
            U-FVA\or U-CDA\or U-TVDA\or M-OL-FVA\or M-OL-CDA\or M-OL-TVDA\or M-IL-FVA\or M-IL-CDA\or M-IL-TVDA
        \fi};
}
\draw[blue, thick] 
    (0:0.54711606*\radius) -- (40:0.645589917*\radius) -- (80:0.669352806*\radius) --
    (120:0.516555889*\radius) -- (160:0.584763524*\radius) -- (200:0.566789966*\radius) --
    (240:0.514576281*\radius) -- (280:0.623773397*\radius) -- (320:0.582132432*\radius) -- cycle;
\draw[red, thick] 
    (0:0.715288078*\radius) -- (40:0.785546154*\radius) -- (80:0.998262886*\radius) --
    (120:0.404998826*\radius) -- (160:0.432063199*\radius) -- (200:0.434894458*\radius) --
    (240:0.460912997*\radius) -- (280:0.439644551*\radius) -- (320:0.516928836*\radius) -- cycle;
\draw[orange, thick] 
    (0:0.75*\radius) -- (40:0.787*\radius) -- (80:0.883894573*\radius) --
    (120:0.989948723*\radius) -- (160:0.97652697*\radius) -- (200:0.993647131*\radius) --
    (240:0.935326851*\radius) -- (280:0.982097287*\radius) -- (320:0.984059483*\radius) -- cycle;
\begin{scope}[shift={(0,-3cm)}]
    \draw[blue, thick] (-3,0) -- (-2.5,0) node[right, font=\scriptsize] {DeepSeek-V3};
    \draw[red, thick] (-0.5,0) -- (0,0) node[right, font=\scriptsize] {Qwen3};
    \draw[orange, thick] (1.5,0) -- (2,0) node[right, font=\scriptsize] {GCAD};
\end{scope}
\end{tikzpicture}
\caption{AUPRC}
\label{fig:radar2}
\end{subfigure}

\label{fig:main}
\caption{RQ1--Radar plots of AUROC and AUPRC scores across nine tasks for \Predictionbased approaches}
\label{fig:auroc and auprc}
\end{figure*}
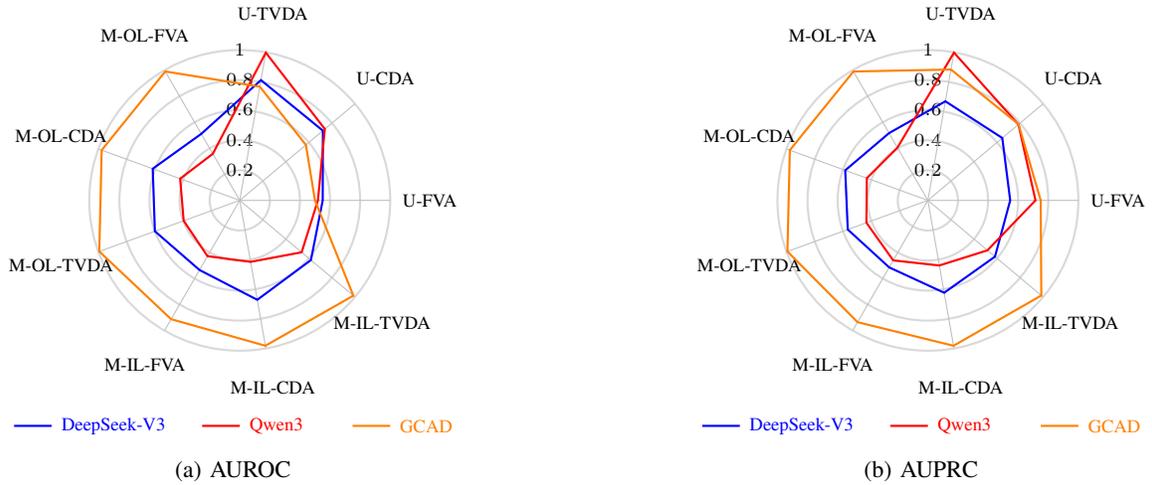

\subsection{Experiments Setups}
\subsubsection{Hyperparameter Configurations.}
Both unsupervised learning approaches and LLMs rely on the sliding window strategy as discussed in \cref{sec:Unsupervised Anomaly Detection} and \cref{sec:Zero-Shot Anomaly Detection with LLMs}, making window-related hyperparameters critical to performance. For unsupervised learning approaches, we followed the recommended configurations from their original publications and reported results determined according to the best F1-score. For fair comparison with LLMs, we conducted preliminary experiments to identify optimal window and step sizes (details in the online supplementary materials\footnote{\url{https://github.com/liuyang2001/ATSADBench}\label{fn:github}}, and adopted the configuration yielding the best results. All window settings are summarized in Table~\ref{tab:window_configs}. All LLMs used greedy decoding (temperature = 0) to ensure stable, reproducible outputs. 

\subsubsection{Enhancement Strategies for LLMs}
To further assess LLM performance in anomaly detection, we explored two strategies: few-shot learning and RAG, both designed to better integrate domain knowledge into the detection process. The detailed prompt structures are provided in the online supplementary materials\footnote{\url{https://github.com/liuyang2001/ATSADBench}\label{fn:github}}.

\textbf{Few-shot Learning Setting.}
The prompt design was tailored to each detection paradigm. For the \Directly paradigm, the prompt included one positive example (a normal temporal sequence manually selected from the training set) and three negative examples, each representing a synthetic anomaly type with accompanying explanations. For the \Predictionbased paradigm, only the positive example was included in the prompt to avoid biasing the model’s predictions. 

\textbf{RAG Setting.}
In the RAG setting, reference windows were retrieved from the training set based on Euclidean distance~\cite{DBLP:conf/icassp/YangWZJ25}. The most similar window was incorporated into the LLM prompt as additional context.

All experiments were conducted on a workstation equipped with an Intel Xeon Platinum 8358P (128 cores), 503 GB RAM, and NVIDIA A100 (80 GB) GPUs.

\section{Results and Analysis}
\subsection{Evaluation of RQ1}
We compare the performance of LLMs (DeepSeek-V3 and Qwen3) against SOTA unsupervised learning approaches (GCAD, TFMAE, Sub-Adjacent) across U, M-OL, and M-IL scenarios. Evaluation is conducted using F1-score, AUROC, and AUPRC as primary metrics. Table~\ref{tab:RQ1 performance} and Fig.~\ref{fig:auroc and auprc} present the experimental results on \benchmark.

\textbf{(1) Zero-shot LLMs vs. Unsupervised Learning Approaches:}
LLMs exhibit inferior average performance compared to unsupervised approaches. Under the \textit{\Directly} paradigm, the best LLM (DeepSeek-V3) achieves an average F1-score of 0.4694, which is 10.46\% lower than the top-performing unsupervised approach, TFMAE (See Table~\ref{tab:RQ1 performance}). In the \textit{\Predictionbased} paradigm, LLMs achieve a maximum average F1-score of 0.6127 (DeepSeek-V3), trailing GCAD by 11.11\% (See Table~\ref{tab:RQ1 performance}). The AUROC and AUPRC scores of LLMs are also significantly lower than those of GCAD (See Fig.~\ref{fig:auroc and auprc}). This performance gap largely stems from the specialized domain knowledge required for aerospace TSAD, highlighting the limitations of generic, off-the-shelf LLMs relative to domain-trained models.

\textbf{(2) FVA vs. CDA vs. TVDA:}
The performance of LLMs varies across the three anomaly types, with no single type proving consistently easier to detect. This variability is dependent on the chosen paradigm, signal dimensionality, and the specific model. For instance, in the univariate tasks, DeepSeek-V3 under the \Directly paradigm achieves its best F1-score on TVDA (0.6199), whereas Qwen3 performs best on FVA (0.7760) (See Table~\ref{tab:RQ1 performance}). Besides, Qwen3 performs best on FVA in the \Directly paradigm but on TVDA in the \Predictionbased paradigm (See Table~\ref{tab:RQ1 performance}). Therefore, we cannot draw a general conclusion about which anomaly type is inherently more or less challenging for LLMs, as the detection performance is highly context-dependent.

\textbf{(3) Univariate vs. Multivariate (In-Loop and Out-of-Loop):}
LLMs perform best in univariate tasks, sometimes even surpassing SOTA unsupervised approaches, while their effectiveness declines as task complexity increases. In M tasks, their overall performance on M-IL tasks is slightly higher than on M-OL tasks. For univariate tasks, Qwen3 under the \textit{\Predictionbased} paradigm achieves the highest average F1-score (0.7118), outperforming M-IL and M-OL settings by 14.37\% and 15.37\%, respectively (See Table~\ref{tab:RQ1 performance}). The AUROC and AUPRC scores show similar trends (See Fig.~\ref{fig:auroc and auprc}). By analyzing the outputs, we found that the performance drop mainly comes from mistaking normal fluctuations in other variables for anomalies, which weakens the visibility of true anomalies. The overall performance in M-IL tasks is better than in M-OL tasks (by 1–5\% in F1-score (See Table~\ref{tab:RQ1 performance}) and by 2-7\% in both AUROC and AUPRC scores (See Fig.~\ref{fig:auroc and auprc})). This may be attributed to in-loop propagation generating larger and more pronounced anomalous patterns in the data, thereby facilitating improved detection performance by the LLMs.

\textbf{(4) \Directly vs. \Predictionbased Paradigm:}
LLMs under the \textit{\Predictionbased} paradigm achieve F1-scores 15–19\% higher than under the \textit{\Directly} paradigm (see Table~\ref{tab:RQ1 performance}). We attribute this to the inherent auto-regressive strengths of LLMs, which are better exploited in the \Predictionbased setting, while the \Directly paradigm requires more abstract, zero-shot classification, which is a more challenging task.

\begin{tcolorbox}[colback=gray!20,boxrule=0pt,top=1mm,bottom=1mm,left=1mm,right=1mm]
\textbf{Answer to RQ1:} 
LLMs perform notably well in U tasks, though they still lag behind SOTA unsupervised approaches. In M tasks, they show a slight advantage in M-IL over M-OL. The effectiveness of LLMs varies across anomaly types. Among the two paradigms, LLMs achieve better results under the \Predictionbased paradigm. 
\end{tcolorbox}

\begin{table}[!tb]
\centering
{
\small
\caption{RQ2--Performance on AA, AL, and AC. The best performance in each column is \textbf{bold}.}
\label{tab:alarm_latency_and_contiguity}
\setlength{\tabcolsep}{1.5pt} 
\begin{tabular}{c|ccccccccc}
\toprule
\multirow{2}{*}{{\textbf{LLM}}} & \multicolumn{3}{c}{\textbf{AA}} & \multicolumn{3}{c}{\textbf{AL}}&\multicolumn{3}{c}{\textbf{AC}}\\
\cmidrule(lr){2-4} \cmidrule(lr){5-7} \cmidrule(lr){8-10}
& U & M-OL & M-IL & U & M-OL & M-IL & U & M-OL & M-IL \\
\hline
\multicolumn{10}{c}{\textit{\Directly}} \\
DeepSeek-V3 & 0.53 & 0.49 & 0.49 & 4.17  & 1.83 & \textbf{1.17} & 0.37 & 0.07 & \textbf{0.08} \\
Qwen3 & 0.45 & 0.51 & 0.48 & 2.33 & 1.83 & 1.33 & 0.22 & \textbf{0.09} & 0.07 \\
\multicolumn{10}{c}{\textit{\Predictionbased}} \\
DeepSeek-V3 & \textbf{0.71} & \textbf{0.56} & \textbf{0.57} & \textbf{1.17} & 2.50 & 1.83 & 0.55 & 0.05 & 0.05 \\
Qwen3 & 0.70 & 0.41 & 0.47 & \textbf{1.17} & \textbf{1.17} & 1.33 & \textbf{0.68} &0.04 & 0.07 \\
\bottomrule
\end{tabular}
}
\end{table}

\subsection{Evaluation of RQ2}
We evaluate the practical performance of the SOTA LLMs, DeepSeek-V3 and Qwen3, on three metrics: AA, AL, and AC, as described in~\cref{sec:Evaluation Metrics}. Results are shown in Table~\ref{tab:alarm_latency_and_contiguity}, where higher AA/AC and lower AL indicate better performance.

\textbf{(1) Performance in AA:} 
LLMs perform optimally only in univariate tasks under the \Predictionbased paradigm, achieving an accuracy of nearly 0.7. Conversely, in all other cases, their accuracy drops to a range between 0.41 and 0.57, which is almost indistinguishable from random guessing. This may be due to the inherent limitations of off-the-shelf LLMs in processing inter-variate relationships and their lack of domain-specific knowledge in aerospace.

\textbf{(2) Performance in AL:} Qwen3 under the \Predictionbased paradigm exhibited the best overall alarm latency, raising an alarm within an average of 1.22 anomalous windows. In other cases, however, significant delays can occur, with alarms sometimes not being issued until the fourth window, which could result in immeasurable damage in aerospace software.

\textbf{(3) Performance in AC:} LLMs achieve significantly higher alarm contiguity in univariate tasks compared to multivariate tasks. In the univariate tasks, the best AC score reaches up to 0.68; however, in the multivariate tasks, AC drops sharply, with the highest score being only 0.09. This further highlights the limited ability of LLMs to analyze relationships among multiple variables.

\begin{tcolorbox}[colback=gray!20,boxrule=0pt,top=1mm,bottom=1mm,left=1mm,right=1mm]
\textbf{Answer to RQ2:} LLMs achieve higher AA and AC only in univariate tasks, whereas their performance in multivariate tasks is nearly equivalent to random guessing (AA close to 0.5, AC averaging 0.06). AL does not follow this trend; the best result is achieved by Qwen3 under the \Predictionbased paradigm, with an average of 1.22 windows.
\end{tcolorbox}

\begin{table}[!tb]
\centering
\small
\setlength{\tabcolsep}{9.5pt}
{
\caption{RQ3--Performance comparison of enhancement strategies of LLMs.}
\label{tab:enhancement_summary}
\begin{tabular}{c|cccc}
\toprule
\textbf{Strategy} & \textbf{F1-Score} & \textbf{AA} & \textbf{AL} & \textbf{AC} \\
\hline
\multicolumn{5}{c}{\textit{DeepSeek-V3 with \Directly paradigm}} \\
ZeroShot & 0.4694 & 0.5024 & 2.4 & 0.1706 \\
FewShot & \perfchange{11.25} & \perfchange{-1.49} & \alchange{-0.5} & \perfchange{40.17} \\
RAG & \perfchange{-1.65} & \perfchange{0.60} & \alchange{-0.2} & \perfchange{-2.00} \\
\hline
\multicolumn{5}{c}{\textit{Qwen3 with \Directly paradigm}} \\
ZeroShot & 0.3342 & 0.4804 & 1.8 & 0.1244 \\
FewShot & \perfchange{28.23} & \perfchange{-0.17} & \alchange{-0.2} & \perfchange{60.22} \\
RAG & \perfchange{15.52} & \perfchange{-0.60} & \alchange{0.0} & \perfchange{-2.17} \\
\hline
\multicolumn{5}{c}{\textit{DeepSeek-V3 with \Predictionbased paradigm}} \\
ZeroShot & 0.6127 & 0.6131 & 1.8 & 0.2178 \\
FewShot & \perfchange{-9.40} & \perfchange{-8.93} & \alchange{0.4} & \perfchange{-1.47} \\
RAG & \perfchange{-4.80} & \perfchange{-4.50} & \alchange{0.4} & \perfchange{-1.08} \\
\hline
\multicolumn{5}{c}{\textit{Qwen3 with \Predictionbased paradigm}} \\
ZeroShot & 0.5334 & 0.5254 & 1.2 & 0.2617 \\
FewShot & \perfchange{3.71} & \perfchange{3.81} & \alchange{0.7} & \perfchange{1.28} \\
RAG & \perfchange{2.27} & \perfchange{0.47} & \alchange{0.1} & \perfchange{3.36} \\
\bottomrule
\end{tabular}
}
\end{table}

\subsection{Evaluation of RQ3}
\label{sec:Evaluation of RQ3}

We evaluate the performance of LLMs on \benchmark
using few-shot learning and retrieved normal temporal sequences. We use both standard classification metrics and practical metrics described in~\cref{sec:Evaluation Metrics}. Table~\ref{tab:enhancement_summary} presents the results.  \textcolor{OliveGreen}{Green} text indicates a performance improvement (higher F1/AA/AC, lower AL), and \textcolor{BrickRed}{red} indicates a degradation.

\textbf{(1) Performance of Few-Shot Learning:} Few-shot learning greatly enhances the performance of LLMs under the Direct paradigm, but shows only slight improvements or even substantial declines under the Prediction-Based paradigm. 
Under the Direct paradigm, Few-shot learning significantly improves DeepSeek-V3 by 11.25\% in F1-score and 40.17\% in AC, and Qwen3 by 28.23\% in F1-score and 60.22\% in AC. In contrast, the Prediction-Based paradigm shows only modest improvements (0.4\% to 4\%) for Qwen3, with increased AL, while DeepSeek-V3 sees nearly 10\% declines in F1-score and AA. This may be due to the absence of negative data points in the Prediction-Based paradigm.

\textbf{(2) Performance of RAG:} LLMs with RAG do not show significant improvements.
For the three metrics (F1-score, AA, and AC), changes remain within the -5\% to +4\% range, except for Qwen3 with the \Directly paradigm, which shows a notable +15.52\% improvement in F1-score. And the changes of AL fluctuate within -0.2 to +0.4. This may be due to the inability of LLMs to extract domain-specific knowledge from retrieved normal temporal sequences.

\begin{tcolorbox}[colback=gray!20,boxrule=0pt,top=1mm,bottom=1mm,left=1mm,right=1mm]
\textbf{Answer to RQ3:} Few-shot learning enhances the performance of LLMs under the \Directly paradigm but exhibits limited capacity under the Prediction-Based paradigm. RAG, on the other hand, does not demonstrate significant improvements across the tasks.
\end{tcolorbox}

\captionsetup[subfigure]{skip=0pt} 
\begin{figure}[!tb]
    \centering
    \begin{subfigure}[t]{0.45\textwidth}
        \centering
        \includegraphics[width=\textwidth]{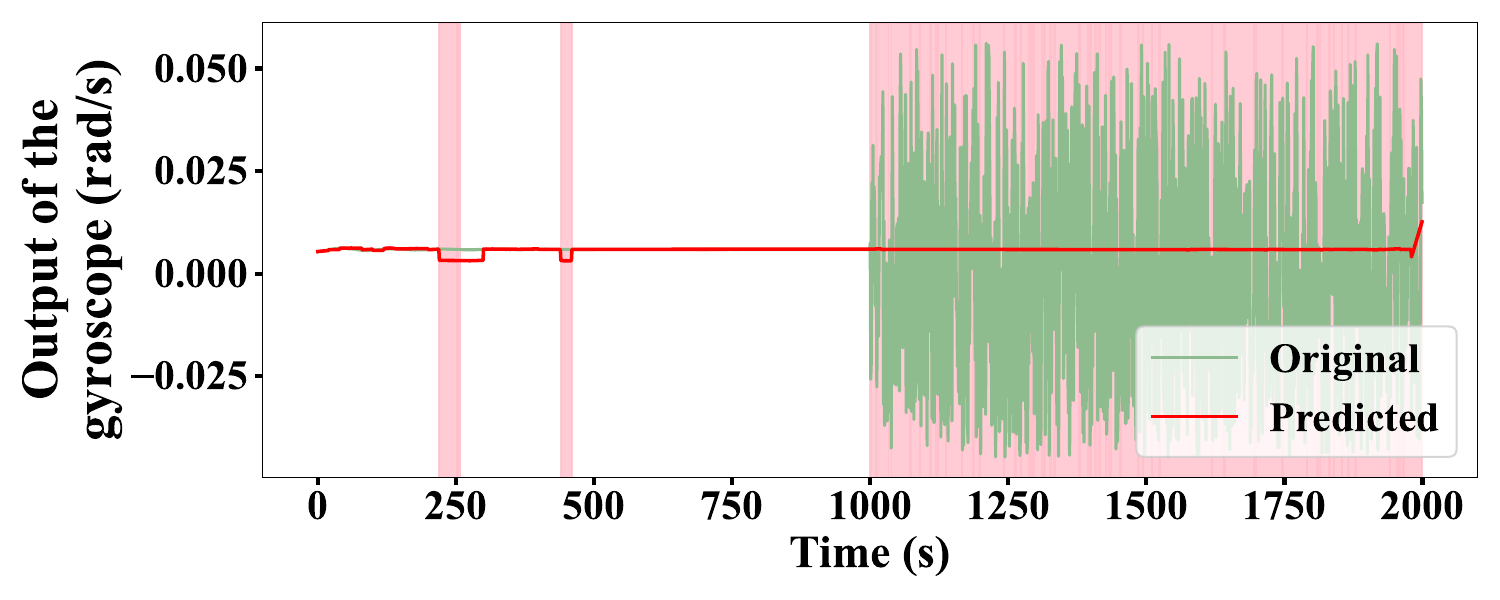}
        \caption{The result of Qwen3 under zero-shot setting.}
        \label{fig:zeroshot}
    \end{subfigure}
    \vfill
    \begin{subfigure}[t]{0.45\textwidth}
        \centering
        \includegraphics[width=\textwidth]{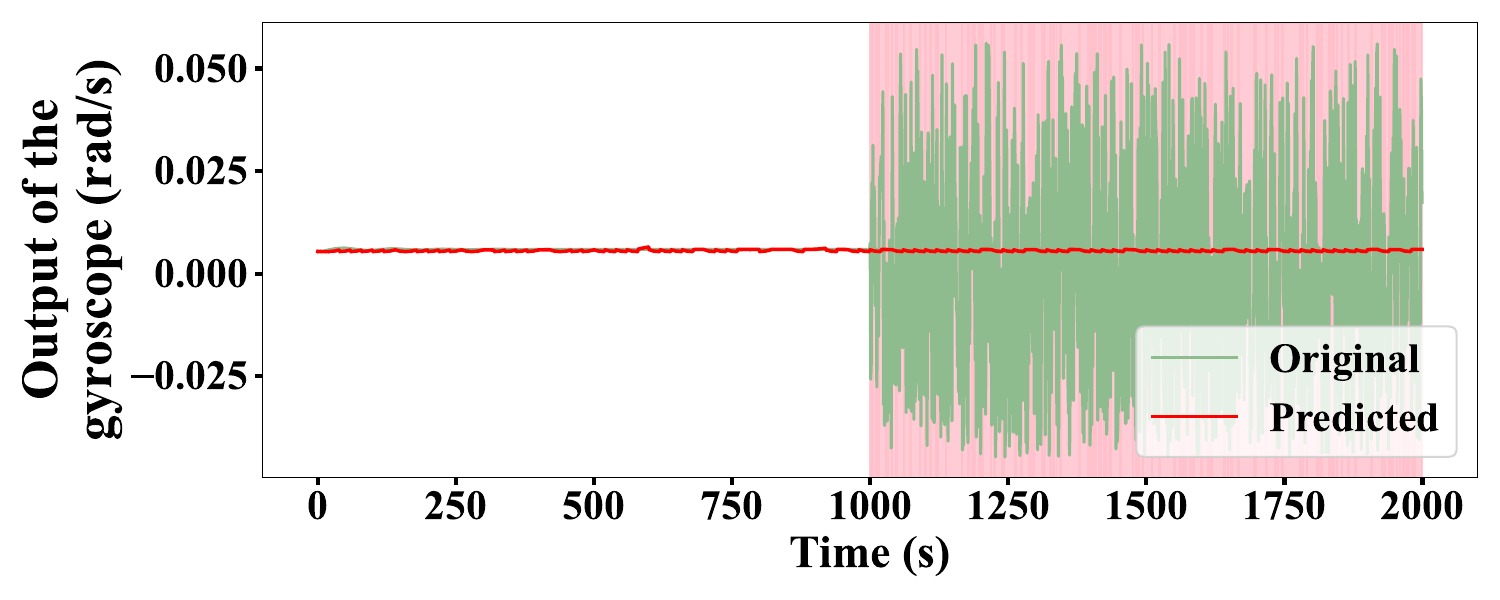}
        \caption{The result of Qwen3 under few-shot setting.}
        \label{fig:fewshot}
    \end{subfigure}
    \vfill
    \begin{subfigure}[t]{0.45\textwidth}
        \centering
        \includegraphics[width=\textwidth]{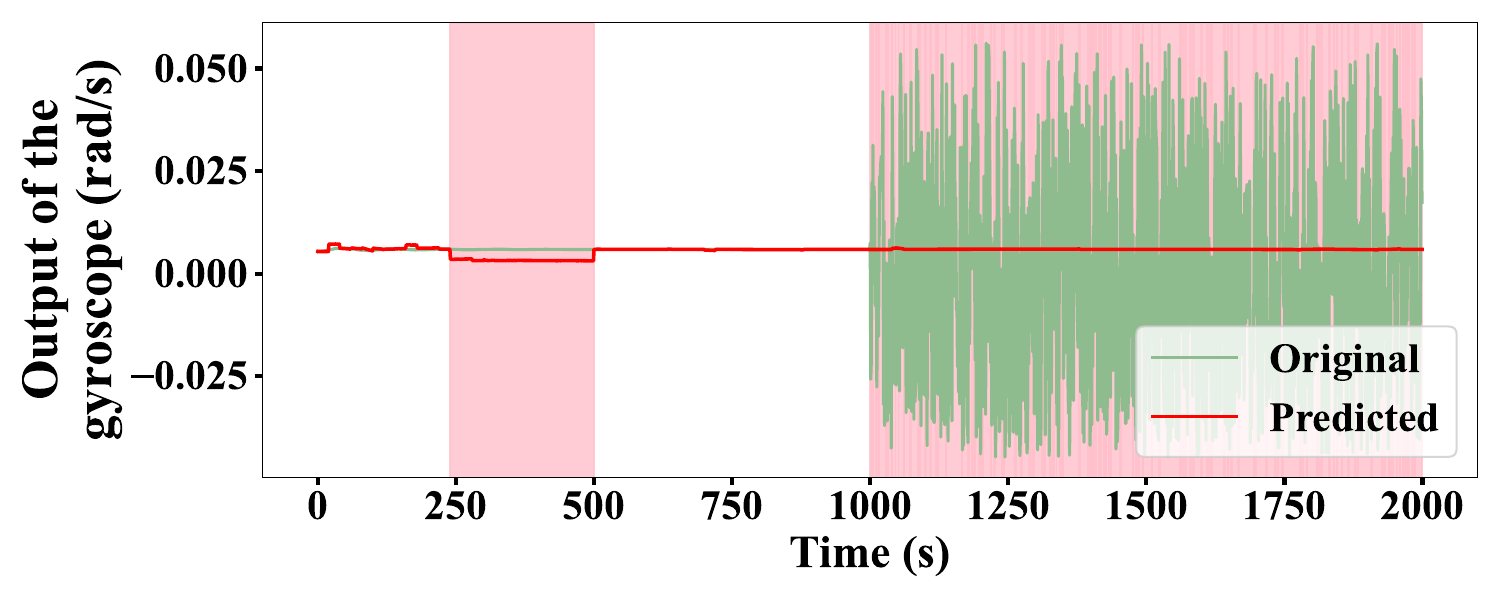}
        \caption{The result of Qwen3 under RAG setting.}
        \label{fig:rag}
    \end{subfigure}
    \hfill
    \caption{RQ4--Case study results of Qwen3 on the gyroscope anomaly under different settings. }
    \label{fig:RQ4}
\end{figure}

\subsection{Evaluation of RQ4}
To validate the practical applicability of LLMs, we conducted a case study on a historical in-orbit anomaly event. Based on our findings in \cref{sec:Evaluation of RQ3}, we selected the Qwen3 model under the \Predictionbased paradigm and examined its performance under zero-shot, few-shot, and RAG settings.

\textbf{Case Description.} We utilized 45-dimensional telemetry data from a significant satellite anomaly on July 12, 2022. The event was initiated by a gyroscope sensor fault that, due to being in-loop, propagated through the attitude control system, ultimately causing structural damage. For our test, we extracted a 2000-second data segment, where the anomaly onset occurs at the 1000-second mark, as illustrated in Fig.~\ref{fig:RQ4}.

\textbf{Results and Analysis.}
The results for the zero-shot, few-shot, and RAG settings are presented in Fig~\ref{fig:zeroshot},~\ref{fig:fewshot} and~\ref{fig:rag}. In each subfigure, the green line represents the original gyroscope output, the red line shows the Qwen3 prediction, and the pink shaded areas indicate data points that the model flagged as anomalous.
A key observation across all three settings is that they successfully generate timely and continuous alarms for the actual anomaly that begins at the 1000s mark. In the zero-shot setting (Fig.\ref{fig:RQ4}a), Qwen3 also produces false alarms in the intervals [220s, 257s] and [440s, 459s]. By contrast, the few-shot setting (Fig.\ref{fig:RQ4}b) eliminates these false alarms and achieves an F1-score of 0.9379, AA of 0.95, AL of 1, and AC of 0.52. The RAG setting (Fig.\ref{fig:RQ4}c), however, introduces a longer false alarm period from 240s to 499s.
\begin{tcolorbox}[colback=gray!20,boxrule=0pt,top=1mm,bottom=1mm,left=1mm,right=1mm]
\textbf{Answer to RQ4:} In the zero-shot setting, Qwen3 successfully detects the true anomaly onset but produces false alarms prior to the event. Few-shot learning eliminates these false alarms, while RAG leads to more false alarms.
\end{tcolorbox}

\subsection{Threats to Validity}
The threats to internal validity lie in the construction of dataset, as \benchmark is synthetically generated and may not fully capture the complexity of real aerospace anomalies. To address this, we employed high-fidelity simulation, incorporated a wide range of anomaly types and scenarios, and applied rigorous preprocessing to enhance data quality. For unsupervised learning approaches, we adopted recommended hyperparameter configurations and reported the best results. 
The threats to external validity are the generalizability of our findings beyond the constructed benchmark and selected models. Our evaluation is limited to the benchmark we constructed and a subset of open-source LLMs, and results may not generalize to other architectures, proprietary models or datasets. To encourage broader validation, we have released our code\footref{fn:github}, and invite future studies to evaluate additional models and more diverse real-world datasets.

\section{Discussion}

\subsection{Implications on Practice} 
The results from this study offer meaningful insights into the practical deployment of LLM-based anomaly detection within the aerospace sector, for both ground and onboard telemetry systems. Aerospace users can leverage the generalizable nature of LLMs to implement flexible, training-free anomaly detectors, enabling rapid deployment and seamless adaptation to evolving operational telemetry streams. 

\subsection{Limitations} 
We acknowledge several limitations. First, our benchmark is built from simulation data, and the injected anomalies may not fully represent the complexity and diversity of faults encountered in real-world aerospace operations. To enhance authenticity, we plan to develop future benchmarks based on actual satellite telemetry. Second, this work focuses exclusively on the aerospace domain. Applying our evaluation framework to other fields would help assess the generalizability and comparative performance of LLMs for TSAD across different domains. Third, the scope of model selection and enhancement strategies in this work remains limited. We only experimented with a small set of open-source models (DeepSeek and Qwen) and employed relatively simple strategies such as few-shot learning and RAG. Future work will explore a broader range of foundation models, including LLaMA and Mistral, and incorporate more advanced techniques such as reinforcement learning from human feedback (RLHF) and advanced contextual prompts~\cite{DBLP:conf/icassp/LiZKQX25} to further improve detection performance.
\subsection{Future Directions}

\textbf{Enhancing Inter-Variable Reasoning in LLMs.} 
A major challenge is that the natural fluctuations of other normal variables introduce noise, leading LLMs to misinterpret them as anomalies. Our results show that, without domain adaptation, LLM performance drops sharply on multivariate data as additional sensor streams introduce noise and obscure true anomalies. Future work should develop structured prompting techniques that explicitly encode inter-variable relationships such as causal links or physical constraints~\cite{DBLP:conf/iclr/ShojaeeMGFR25,DBLP:journals/telo/MeyersonNBGKHL24,DBLP:journals/nature/RomeraParedesBNBKDREWFKF24,wang2025drsr}, so the model can better distinguish meaningful patterns from spurious correlations in complex telemetry.

\textbf{Developing Practical Metrics.} 
Our work underscores that traditional metrics like F1 score are insufficient for assessing an anomaly detector’s utility in safety-critical aerospace contexts. In such settings, when and how alarms are raised can be as important as overall accuracy. Industrial deployment demands a comprehensive evaluation framework that accounts for alarm timeliness, credibility, false-alarm rates, and operational constraints like model size and inference latency. Future work should focus on designing and standardizing domain-specific metrics that capture these practical requirements and trade-offs~\cite{puccetti2024detection}, supporting more informed model selection and deployment in industrial settings.

\textbf{Integrating Domain Knowledge into LLMs.}
Our results indicate that prompt-based enhancement strategies alone are insufficient for LLMs to effectively handle complex aerospace telemetry, particularly in \Predictionbased anomaly detection. Achieving robust domain adaptation requires more fundamental approaches, such as domain-specific pretraining and fine-tuning~\cite{DBLP:journals/corr/abs-2401-02981,DBLP:journals/corr/abs-2402-15061,DBLP:conf/icml/ShenTHAF24}. By training LLMs on large corpora of in-domain time-series data and embedding relevant engineering knowledge, these approaches enable models to internalize essential physical relationships and operational patterns. This integration, which goes beyond zero-shot learning, is essential for narrowing the performance gap with specialized unsupervised learning approaches and enhancing detection accuracy and credibility in real-world aerospace applications.

\textbf{Generality to Other Domains.}
Current TSAD research predominantly evaluates approaches on generic datasets spanning multiple domains, which may mask domain-specific challenges and limit the practical relevance of reported results. In contrast, our study demonstrates the benefits of a focused, in-depth evaluation in aerospace, leveraging domain-specific data and requirements. We encourage researchers in other critical application areas to conduct similarly targeted investigations, developing dedicated benchmarks, capturing the distinct characteristics of their domains, and designing practical evaluation metrics that align with real-world user needs.

\section{CONCLUSION}
In this paper, we conducted the first systematic evaluation of LLMs for TSAD in the aerospace domain. To support this study, we introduce \benchmark, a benchmark constructed from realistic aerospace scenarios with 108,000 data points, and propose three user-oriented, window-based metrics: Alarm Accuracy, Alarm Latency, and Alarm Contiguity. These metrics are designed to provide a practical and comprehensive assessment of model performance in operational contexts.

Our experimental results show that LLMs achieve promising results in univariate TSAD tasks but encounter significant difficulties with multivariate telemetry, where their performance is often close to random guessing. Additionally, our study finds that enhancement strategies offer only limited benefits; few-shot learning brings moderate improvements under the \Directly paradigm, while RAG does not lead to substantial gains. We hope that the introduction of \benchmark and user-oriented metrics will inspire further research into LLMs for TSAD and support the development of more effective approaches for aerospace and other safety-critical domains.


\section*{Acknowledgment}
This work is supported by the National Natural Science Foundation of China (Grant Nos. 62402038, U21B2015, 62192730 and 62192735).

\bibliographystyle{IEEEtran}
\bibliography{main}

@inproceedings{he2022share,
  title={Share or not share? towards the practicability of deep models for unsupervised anomaly detection in modern online systems},
  author={He, Zilong and Chen, Pengfei and Huang, Tao},
  booktitle={Proc. IEEE 33rd International Symposium on Software Reliability Engineering (ISSRE)},
  pages={25--35},
  year={2022},
  organization={IEEE}
}

@article{wu2021current,
  title={Current time series anomaly detection benchmarks are flawed and are creating the illusion of progress},
  author={Wu, Renjie and Keogh, Eamonn J},
  journal={IEEE transactions on knowledge and data engineering},
  volume={35},
  number={3},
  pages={2421--2429},
  year={2021},
  publisher={IEEE}
}

@article{jin2024survey,
  title={A survey on graph neural networks for time series: Forecasting, classification, imputation, and anomaly detection},
  author={Jin, Ming and Koh, Huan Yee and Wen, Qingsong and Zambon, Daniele and Alippi, Cesare and Webb, Geoffrey I and King, Irwin and Pan, Shirui},
  journal={IEEE Transactions on Pattern Analysis and Machine Intelligence},
  year={2024},
  publisher={IEEE}
}

@inproceedings{puccetti2024detection,
  title={Detection Latencies of Anomaly Detectors-An Overlooked Perspective?},
  author={Puccetti, Tommaso and Ceccarelli, Andrea},
  booktitle={IEEE 35th International Symposium on Software Reliability Engineering (ISSRE)},
  pages={37--48},
  year={2024},
  organization={IEEE}
}

@inproceedings{DBLP:conf/ijcai/FreemanMB022,
  author       = {Cynthia Freeman and
                  Jonathan Merriman and
                  Ian Beaver and
                  Abdullah Mueen},
  editor       = {Luc De Raedt},
  title        = {Experimental Comparison and Survey of Twelve Time Series Anomaly Detection Algorithms (Extended Abstract)},
  booktitle    = {Proceedings of the Thirty-First International Joint Conference on
                  Artificial Intelligence, {IJCAI}},
  pages        = {5737--5741},
  publisher    = {ijcai.org},
  year         = {2022}
}

@article{sarfraz2024position,
  title={Position: quo vadis, unsupervised time series anomaly detection?},
  author={Sarfraz, M Saquib and Chen, Mei-Yen and Layer, Lukas and Peng, Kunyu and Koulakis, Marios},
  journal={arXiv preprint arXiv:2405.02678},
  year={2024}
}

@inproceedings{li2023empirical,
  title={An empirical analysis of anomaly detection methods for multivariate time series},
  author={Li, Dongwen and Zhang, Shenglin and Sun, Yongqian and Guo, Yang and Che, Zeyu and Chen, Shiqi and Zhong, Zhenyu and Liang, Minghan and Shao, Minyi and Li, Mingjie and others},
  booktitle={IEEE 34th International Symposium on Software Reliability Engineering (ISSRE)},
  pages={57--68},
  year={2023},
  organization={IEEE}
}

@inproceedings{kim2022towards,
  title={Towards a rigorous evaluation of time-series anomaly detection},
  author={Kim, Siwon and Choi, Kukjin and Choi, Hyun-Soo and Lee, Byunghan and Yoon, Sungroh},
  booktitle={Proceedings of the AAAI Conference on Artificial Intelligence},
  volume={36},
  number={7},
  pages={7194--7201},
  year={2022}
}

@article{audibert2022deep,
  title={Do deep neural networks contribute to multivariate time series anomaly detection?},
  author={Audibert, Julien and Michiardi, Pietro and Guyard, Fr{\'e}d{\'e}ric and Marti, S{\'e}bastien and Zuluaga, Maria A},
  journal={Pattern Recognition},
  volume={132},
  pages={108945},
  year={2022},
  publisher={Elsevier}
}

@inproceedings{liu2023practical,
  title={Practical anomaly detection over multivariate monitoring metrics for online services},
  author={Liu, Jinyang and Yang, Tianyi and Chen, Zhuangbin and Su, Yuxin and Feng, Cong and Yang, Zengyin and Lyu, Michael R},
  booktitle={IEEE 34th International Symposium on Software Reliability Engineering (ISSRE)},
  pages={36--45},
  year={2023},
  organization={IEEE}
}

@inproceedings{lee2023maat,
  title={Maat: Performance metric anomaly anticipation for cloud services with conditional diffusion},
  author={Lee, Cheryl and Yang, Tianyi and Chen, Zhuangbin and Su, Yuxin and Lyu, Michael R},
  booktitle={Proceedings of the 38th IEEE/ACM International Conference on Automated Software Engineering (ASE)},
  pages={116--128},
  year={2023},
  organization={IEEE}
}

@inproceedings{huang2025graph,
  title={Graph mixture of experts and memory-augmented routers for multivariate time series anomaly detection},
  author={Huang, Xiaoyu and Chen, Weidong and Hu, Bo and Mao, Zhendong},
  booktitle={Proceedings of the AAAI Conference on Artificial Intelligence},
  volume={39},
  number={16},
  pages={17476--17484},
  year={2025}
}

@inproceedings{chen2022deep,
  title={Deep variational graph convolutional recurrent network for multivariate time series anomaly detection},
  author={Chen, Wenchao and Tian, Long and Chen, Bo and Dai, Liang and Duan, Zhibin and Zhou, Mingyuan},
  booktitle={International conference on machine learning},
  pages={3621--3633},
  year={2022},
  organization={PMLR}
}

@article{jing2024diner,
  title={Diner: Interpretable Anomaly Detection for Seasonal Time Series in Web Services},
  author={Jing, Yuhan and Wang, Jingyu and Qi, Ji and Qi, Qi and He, Bo and Zhuang, Zirui and Wu, Naixing and Liao, Jianxin},
  journal={IEEE Transactions on Services Computing},
  year={2024},
  publisher={IEEE}
}

@article{DBLP:journals/tkde/ZhangXCCZXY24,
  author       = {Xiao Zhang and
                  Shuqing Xu and
                  Huashan Chen and
                  Zekai Chen and
                  Fuzhen Zhuang and
                  Hui Xiong and
                  Dongxiao Yu},
  title        = {Rethinking Robust Multivariate Time Series Anomaly Detection: {A}
                  Hierarchical Spatio-Temporal Variational Perspective},
  journal      = {{IEEE} Trans. Knowl. Data Eng.},
  volume       = {36},
  number       = {12},
  pages        = {9136--9149},
  year         = {2024},
  doi          = {10.1109/TKDE.2024.3466291},
}

@article{DBLP:journals/tkde/KimKML24,
  author       = {HyunGi Kim and
                  Siwon Kim and
                  Seonwoo Min and
                  Byunghan Lee},
  title        = {Contrastive Time-Series Anomaly Detection},
  journal      = {{IEEE} Trans. Knowl. Data Eng.},
  volume       = {36},
  number       = {10},
  pages        = {5053--5065},
  year         = {2024},
  doi          = {10.1109/TKDE.2023.3335317},
}

@article{DBLP:journals/tkde/XuWJLWP24,
  author       = {Hongzuo Xu and
                  Yijie Wang and
                  Songlei Jian and
                  Qing Liao and
                  Yongjun Wang and
                  Guansong Pang},
  title        = {Calibrated One-Class Classification for Unsupervised Time Series Anomaly
                  Detection},
  journal      = {{IEEE} Trans. Knowl. Data Eng.},
  volume       = {36},
  number       = {11},
  pages        = {5723--5736},
  year         = {2024},
  doi          = {10.1109/TKDE.2024.3393996},
}

@article{DBLP:journals/tkde/ZhengMZZZYYZJ24,
  author       = {Bolong Zheng and
                  Lingfeng Ming and
                  Kai Zeng and
                  Mengtao Zhou and
                  Xinyong Zhang and
                  Tao Ye and
                  Bin Yang and
                  Xiaofang Zhou and
                  Christian S. Jensen},
  title        = {Adversarial Graph Neural Network for Multivariate Time Series Anomaly
                  Detection},
  journal      = {{IEEE} Trans. Knowl. Data Eng.},
  volume       = {36},
  number       = {12},
  pages        = {7612--7626},
  year         = {2024},
  doi          = {10.1109/TKDE.2024.3419891},
}

@article{liu2024uac,
  title={Uac-ad: Unsupervised adversarial contrastive learning for anomaly detection on multi-modal data in microservice systems},
  author={Liu, Hongyi and Huang, Xiaosong and Jia, Mengxi and Jia, Tong and Han, Jing and Li, Ying and Wu, Zhonghai},
  journal={IEEE Transactions on Services Computing},
  year={2024},
  publisher={IEEE}
}

@incollection{li2023deepdiscord,
  title={DeepDiscord: Dual Contrastive Coding for Transferable Time Series Anomaly Detection},
  author={Li, Xin-Yi and Zhong, Pei-Nan and Chen, Di and Zhang, Zhen-Dong and Yang, Yu-Bin},
  booktitle={ECAI 2023},
  pages={1439--1446},
  year={2023},
  publisher={IOS Press}
}

@inproceedings{DBLP:conf/ijcai/ZhangZT22,
  author       = {Weiqi Zhang and
                  Chen Zhang and
                  Fugee Tsung},
  title        = {{GRELEN:} Multivariate Time Series Anomaly Detection from the Perspective
                  of Graph Relational Learning},
  booktitle    = {Proceedings of the Thirty-First International Joint Conference on
                  Artificial Intelligence, {IJCAI}},
  pages        = {2390--2397},
  publisher    = {ijcai.org},
  year         = {2022},
  doi          = {10.24963/IJCAI.2022/332},
}

@inproceedings{zhang2023fkpiselect,
  title={fKPISelect: Fault-Injection Based Automated KPI Selection for Practical Multivariate Anomaly Detection},
  author={Zhang, Xingjian and Zhao, Yinqin and Liu, Chang and Wang, Long and Yang, Xin and Hou, Yefei and Lan, Zhongwen and Hu, Xining and Miao, Beibei and Yang, Ming and others},
  booktitle={IEEE 34th International Symposium on Software Reliability Engineering (ISSRE)},
  pages={183--194},
  year={2023}
}

@inproceedings{li2023tadl,
  title={Tadl: Fault localization with transformer-based anomaly detection for dynamic microservice systems},
  author={Li, Yuewei and Lu, Yan and Wang, Jingyu and Qi, Qi and Wang, Jing and Wang, Yingying and Liao, Jianxin},
  booktitle={IEEE International Conference on Software Analysis, Evolution and Reengineering (SANER)},
  pages={718--722},
  year={2023}
}

@article{tang2024mlp,
  title={MLP-Mixer based Masked Autoencoders are Effective, Explainable and Robust for Time Series Anomaly Detection},
  author={Tang, Qideng and Dai, Chaofan and Wu, Yahui and Zhou, Haohao},
  journal={Proceedings of the VLDB Endowment},
  volume={18},
  number={3},
  pages={798--811},
  year={2024}
}

@inproceedings{wang2024revisiting,
  title={Revisiting vae for unsupervised time series anomaly detection: A frequency perspective},
  author={Wang, Zexin and Pei, Changhua and Ma, Minghua and Wang, Xin and Li, Zhihan and Pei, Dan and Rajmohan, Saravan and Zhang, Dongmei and Lin, Qingwei and Zhang, Haiming and others},
  booktitle={Proceedings of the ACM Web Conference},
  pages={3096--3105},
  year={2024}
}

@inproceedings{kim2024model,
  title={When model meets new normals: test-time adaptation for unsupervised time-series anomaly detection},
  author={Kim, Dongmin and Park, Sunghyun and Choo, Jaegul},
  booktitle={Proceedings of the AAAI conference on artificial intelligence},
  volume={38},
  number={12},
  pages={13113--13121},
  year={2024}
}

@inproceedings{liu2025gcad,
  author       = {Zehao Liu and
                  Mengzhou Gao and
                  Pengfei Jiao},
  title        = {{GCAD:} Anomaly Detection in Multivariate Time Series from the Perspective
                  of Granger Causality},
  booktitle    = {Proceedings of the 39th AAAI Conference on Artificial Intelligence},
  pages        = {19041--19049},
  year         = {2025}
}

@inproceedings{chen2022adaptive,
  title={Adaptive performance anomaly detection for online service systems via pattern sketching},
  author={Chen, Zhuangbin and Liu, Jinyang and Su, Yuxin and Zhang, Hongyu and Ling, Xiao and Yang, Yongqiang and Lyu, Michael R},
  booktitle={Proceedings of the 44th international conference on software engineering},
  pages={61--72},
  year={2022}
}

@inproceedings{zhang2024dyntrackr,
  title={DynTrackr: A Robust Two-Stage Framework with Attribute Enhancement for KPI Anomaly Detection},
  author={Zhang, Meixian and Shi, Xue and Huang, Jiaxin and Su, Lide and Zhang, Yanan},
  booktitle={IEEE 35th International Symposium on Software Reliability Engineering Workshops (ISSREW)},
  pages={169--176},
  year={2024},
  organization={IEEE}
}

@inproceedings{si2023beyond,
  title={Beyond sharing: Conflict-aware multivariate time series anomaly detection},
  author={Si, Haotian and Pei, Changhua and Li, Zhihan and Zhao, Yadong and Li, Jingjing and Zhang, Haiming and Diao, Zulong and Li, Jianhui and Xie, Gaogang and Pei, Dan},
  booktitle={Proceedings of the 31st ACM Joint European Software Engineering Conference and Symposium on the Foundations of Software Engineering},
  pages={1635--1645},
  year={2023}
}

@article{alnegheimish2024large,
  author       = {Sarah Alnegheimish and
                  Linh Nguyen and
                  Laure Berti{-}{\'{E}}quille and
                  Kalyan Veeramachaneni},
  title        = {Large language models can be zero-shot anomaly detectors for time
                  series?},
  journal      = {CoRR},
  volume       = {abs/2405.14755},
  year         = {2024},
  doi          = {10.48550/ARXIV.2405.14755}
}

@article{DBLP:journals/corr/abs-2305-14201,
  author       = {Tiedong Liu and
                  Bryan Kian Hsiang Low},
  title        = {Goat: Fine-tuned LLaMA Outperforms {GPT-4} on Arithmetic Tasks},
  journal      = {CoRR},
  volume       = {abs/2305.14201},
  year         = {2023},
  doi          = {10.48550/ARXIV.2305.14201}
}

@article{DBLP:journals/corr/abs-2307-09288,
  title        = {Llama 2: Open Foundation and Fine-Tuned Chat Models},
  journal      = {CoRR},
  volume       = {abs/2307.09288},
  year         = {2023},
  doi          = {10.48550/ARXIV.2307.09288}
}

@article{DBLP:journals/corr/abs-2505-09388,
  title        = {Qwen3 Technical Report},
  journal      = {CoRR},
  volume       = {abs/2505.09388},
  year         = {2025},
  doi          = {10.48550/ARXIV.2505.09388}
}

@article{DBLP:journals/corr/abs-2412-19437,
  title        = {DeepSeek-V3 Technical Report},
  journal      = {CoRR},
  volume       = {abs/2412.19437},
  year         = {2024},
  doi          = {10.48550/ARXIV.2412.19437},
}

@article{yue2024sub,
  author       = {Wenzhen Yue and
                  Xianghua Ying and
                  Ruohao Guo and
                  DongDong Chen and
                  Ji Shi and
                  Bowei Xing and
                  Yuqing Zhu and
                  Taiyan Chen},
  title        = {Sub-Adjacent Transformer: Improving Time Series Anomaly Detection
                  with Reconstruction Error from Sub-Adjacent Neighborhoods},
  journal      = {CoRR},
  volume       = {abs/2404.18948},
  year         = {2024},
  doi          = {10.48550/ARXIV.2404.18948}
}

@inproceedings{fang2024temporal,
  title={Temporal-frequency masked autoencoders for time series anomaly detection},
  author={Fang, Yuchen and Xie, Jiandong and Zhao, Yan and Chen, Lu and Gao, Yunjun and Zheng, Kai},
  booktitle={IEEE 40th International Conference on Data Engineering (ICDE)},
  pages={1228--1241},
  year={2024},
  organization={IEEE}
}

@article{DBLP:journals/corr/abs-2405-15370,
  author       = {Jun Liu and
                  Chaoyun Zhang and
                  Jiaxu Qian and
                  Minghua Ma and
                  Si Qin and
                  Chetan Bansal and
                  Qingwei Lin and
                  Saravan Rajmohan and
                  Dongmei Zhang},
  title        = {Large Language Models can Deliver Accurate and Interpretable Time
                  Series Anomaly Detection},
  journal      = {CoRR},
  volume       = {abs/2405.15370},
  year         = {2024},
  doi          = {10.48550/ARXIV.2405.15370}
}

@inproceedings{lai2021revisiting,
  title={Revisiting time series outlier detection: Definitions and benchmarks},
  author={Lai, Kwei-Herng and Zha, Daochen and Xu, Junjie and Zhao, Yue and Wang, Guanchu and Hu, Xia},
  booktitle={Thirty-fifth conference on neural information processing systems datasets and benchmarks track (round 1)},
  year={2021}
}

@inproceedings{si2024timeseriesbench,
  title={Timeseriesbench: An industrial-grade benchmark for time series anomaly detection models},
  author={Si, Haotian and Li, Jianhui and Pei, Changhua and Cui, Hang and Yang, Jingwen and Sun, Yongqian and Zhang, Shenglin and Li, Jingjing and Zhang, Haiming and Han, Jing and others},
  booktitle={IEEE 35th International Symposium on Software Reliability Engineering (ISSRE)},
  pages={61--72},
  year={2024}
}

@article{gruver2023large,
  title={Large language models are zero-shot time series forecasters},
  author={Gruver, Nate and Finzi, Marc and Qiu, Shikai and Wilson, Andrew G},
  journal={Advances in Neural Information Processing Systems},
  volume={36},
  pages={19622--19635},
  year={2023}
}

@INPROCEEDINGS{5633180,
  author={Quan Li and XingShe Zhou and Peng Lin and Shaomin Li},
  booktitle={3rd International Symposium on Systems and Control in Aeronautics and Astronautics}, 
  title={Anomaly detection and fault Diagnosis technology of spacecraft based on telemetry-mining}, 
  year={2010},
  volume={},
  number={},
  pages={233-236},
  doi={10.1109/ISSCAA.2010.5633180}}

@article{https://doi.org/10.1609/aimag.v35i4.2553,
author = {Heras, José Martínez and Donati, Alessandro},
title = {Enhanced Telemetry Monitoring with Novelty Detection},
journal = {AI Magazine},
volume = {35},
number = {4},
pages = {37-46},
doi = {https://doi.org/10.1609/aimag.v35i4.2553},
year = {2014}
}

@INPROCEEDINGS{1659593,
  author={Yairi, T. and Kawahara, Y. and Fujimaki, R. and Sato, Y. and Machida, K.},
  booktitle={Proceedings of the 2nd IEEE International Conference on Space Mission Challenges for Information Technology (SMC-IT'06)}, 
  title={Telemetry-mining: a machine learning approach to anomaly detection and fault diagnosis for space systems}, 
  year={2006},
  volume={},
  number={},
  pages={8 pp.-476},
  doi={10.1109/SMC-IT.2006.79}}

@inproceedings{10.1609/aaai.v38i12.29210,
author = {Kim, Dongmin and Park, Sunghyun and Choo, Jaegul},
title = {When model meets new normals: test-time adaptation for unsupervised time-series anomaly detection},
year = {2024},
isbn = {978-1-57735-887-9},
publisher = {AAAI Press},
doi = {10.1609/aaai.v38i12.29210},
articleno = {1462},
numpages = {9},
}

@inproceedings{DBLP:conf/ecai/LiZCZY23,
  author       = {Xin{-}Yi Li and
                  Pei{-}Nan Zhong and
                  Di Chen and
                  Zhendong Zhang and
                  Yu{-}Bin Yang},
  title        = {DeepDiscord: Dual Contrastive Coding for Transferable Time Series
                  Anomaly Detection},
  series       = {Frontiers in Artificial Intelligence and Applications},
  volume       = {372},
  pages        = {1439--1446},
  publisher    = {{IOS} Press},
  year         = {2023},
  doi          = {10.3233/FAIA230422},
}

@inproceedings{DBLP:conf/ijcai/LiuHZLM24,
  author       = {Chen Liu and
                  Shibo He and
                  Qihang Zhou and
                  Shizhong Li and
                  Wenchao Meng},
  title        = {Large Language Model Guided Knowledge Distillation for Time Series
                  Anomaly Detection},
  booktitle    = {Proceedings of the Thirty-Third International Joint Conference on
                  Artificial Intelligence, {IJCAI}},
  pages        = {2162--2170},
  publisher    = {ijcai.org},
  year         = {2024}
}

@inproceedings{DBLP:conf/nips/WangZQ0W0L23,
  author       = {Chengsen Wang and
                  Zirui Zhuang and
                  Qi Qi and
                  Jingyu Wang and
                  Xingyu Wang and
                  Haifeng Sun and
                  Jianxin Liao},
  title        = {Drift doesn't Matter: Dynamic Decomposition with Diffusion Reconstruction
                  for Unstable Multivariate Time Series Anomaly Detection},
  booktitle    = {Advances in Neural Information Processing Systems (NeurIPS)},
  year         = {2023}
}

@inproceedings{DBLP:conf/nips/SongKOC23,
  author       = {Junho Song and
                  Keonwoo Kim and
                  Jeonglyul Oh and
                  Sungzoon Cho},
  title        = {{MEMTO:} Memory-guided Transformer for Multivariate Time Series Anomaly
                  Detection},
  booktitle    = {Advances in Neural Information Processing Systems 36: Annual Conference
                  on Neural Information Processing Systems (NeurIPS)},
  year         = {2023},
}

@inproceedings{DBLP:conf/nips/LaiSGLB23,
  author       = {Chih{-}Yu Lai and
                  Fan{-}Keng Sun and
                  Zhengqi Gao and
                  Jeffrey H. Lang and
                  Duane S. Boning},
  title        = {Nominality Score Conditioned Time Series Anomaly Detection by Point/Sequential
                  Reconstruction},
  booktitle    = {Advances in Neural Information Processing Systems 36: Annual Conference
                  on Neural Information Processing Systems, {NeurIPS} },
  year         = {2023}
}

@article{DBLP:journals/tkde/LiYWJY23,
  author       = {Longyuan Li and
                  Junchi Yan and
                  Qingsong Wen and
                  Yaohui Jin and
                  Xiaokang Yang},
  title        = {Learning Robust Deep State Space for Unsupervised Anomaly Detection
                  in Contaminated Time-Series},
  journal      = {{IEEE} Trans. Knowl. Data Eng.},
  volume       = {35},
  number       = {6},
  pages        = {6058--6072},
  year         = {2023},
  doi          = {10.1109/TKDE.2022.3171562},
}

@article{DBLP:journals/tkde/ZhangCWP23,
  author       = {Yuxin Zhang and
                  Yiqiang Chen and
                  Jindong Wang and
                  Zhiwen Pan},
  title        = {Unsupervised Deep Anomaly Detection for Multi-Sensor Time-Series Signals},
  journal      = {{IEEE} Trans. Knowl. Data Eng.},
  volume       = {35},
  number       = {2},
  pages        = {2118--2132},
  year         = {2023},
  doi          = {10.1109/TKDE.2021.3102110},
}

@article{DBLP:journals/tkde/LiuZDHZSC23,
  author       = {Shenghua Liu and
                  Bin Zhou and
                  Quan Ding and
                  Bryan Hooi and
                  Zhengbo Zhang and
                  Huawei Shen and
                  Xueqi Cheng},
  title        = {Time Series Anomaly Detection With Adversarial Reconstruction Networks},
  journal      = {{IEEE} Trans. Knowl. Data Eng.},
  volume       = {35},
  number       = {4},
  pages        = {4293--4306},
  year         = {2023},
  doi          = {10.1109/TKDE.2021.3140058},
}

@inproceedings{DBLP:conf/issre/YuPZWLXP23,
  title        = {AutoKAD: Empowering {KPI} Anomaly Detection with Label-Free Deployment},
  booktitle    = {{IEEE} International Symposium on Software Reliability Engineering,
                  {ISSRE}},
  pages        = {13--23},
  publisher    = {{IEEE}},
  year         = {2023},
  doi          = {10.1109/ISSRE59848.2023.00063},
}

@inproceedings{DBLP:conf/issre/SunGLWKZLXP24,
  title        = {Multivariate Time Series Anomaly Detection based on Pre-trained Models
                  with Dual-Attention Mechanism},
  booktitle    = {{IEEE} International Symposium on Software Reliability Engineering,
                  {ISSRE}},
  pages        = {73--78},
  publisher    = {{IEEE}},
  year         = {2024},
  doi          = {10.1109/ISSREW63542.2024.00050},
}

@inproceedings{DBLP:conf/icassp/YangWZJ25,
  author       = {Silin Yang and
                  Dong Wang and
                  Haoqi Zheng and
                  Ruochun Jin},
  title        = {TimeRAG: Boosting {LLM} Time Series Forecasting via Retrieval-Augmented
                  Generation},
  booktitle    = {{IEEE} International Conference on Acoustics, Speech and Signal
                  Processing, {ICASSP}},
  pages        = {1--5},
  publisher    = {{IEEE}},
  year         = {2025},
  doi          = {10.1109/ICASSP49660.2025.10889933},
}

@article{DBLP:journals/corr/abs-2305-00118,
  author       = {Kent K. Chang and
                  Mackenzie Cramer and
                  Sandeep Soni and
                  David Bamman},
  title        = {Speak, Memory: An Archaeology of Books Known to ChatGPT/GPT-4},
  journal      = {CoRR},
  volume       = {abs/2305.00118},
  year         = {2023},
  doi          = {10.48550/ARXIV.2305.00118},
}

@inproceedings{DBLP:conf/icde/Liu0C0WWZ23,
  author       = {Fan Liu and
                  Xingshe Zhou and
                  Jinli Cao and
                  Zhu Wang and
                  Tianben Wang and
                  Hua Wang and
                  Yanchun Zhang},
  title        = {Anomaly Detection in Quasi-Periodic Time Series based on Automatic
                  Data Segmentation and Attentional {LSTM-CNN} (Extended Abstract)},
  booktitle    = {Proceedings of the 39th {IEEE} International Conference on Data Engineering, {ICDE}},
  pages        = {3777--3778},
  publisher    = {{IEEE}},
  year         = {2023},
  doi          = {10.1109/ICDE55515.2023.00315},
}

@inproceedings{DBLP:conf/cloud/ZimmermanKSCFBK19,
  author       = {Zachary Zimmerman and
                  Kaveh Kamgar and
                  Nader Shakibay Senobari and
                  Brian Crites and
                  Gareth J. Funning and
                  Philip Brisk and
                  Eamonn J. Keogh},
  title        = {Matrix Profile {XIV:} Scaling Time Series Motif Discovery with GPUs
                  to Break a Quintillion Pairwise Comparisons a Day and Beyond},
  booktitle    = {Proceedings of the {ACM} Symposium on Cloud Computing (SoCC)},
  pages        = {74--86},
  publisher    = {{ACM}},
  year         = {2019},
  doi          = {10.1145/3357223.3362721},
}

@inproceedings{DBLP:conf/kdd/PangSH19,
  author       = {Guansong Pang and
                  Chunhua Shen and
                  Anton van den Hengel},
  title        = {Deep Anomaly Detection with Deviation Networks},
  booktitle    = {Proceedings of the 25th {ACM} {SIGKDD} International Conference on
                  Knowledge Discovery {\&} Data Mining (KDD)},
  pages        = {353--362},
  publisher    = {{ACM}},
  year         = {2019},
  doi          = {10.1145/3292500.3330871},
}

@article{DBLP:journals/symmetry/QiuDQ19,
  author       = {Juan Qiu and
                  Qingfeng Du and
                  Chongshu Qian},
  title        = {{KPI-TSAD:} {A} Time-Series Anomaly Detector for {KPI} Monitoring
                  in Cloud Applications},
  journal      = {Symmetry},
  volume       = {11},
  number       = {11},
  pages        = {1350},
  year         = {2019},
  doi          = {10.3390/SYM11111350},
}

@article{DBLP:journals/kbs/Villa-PerezCLMV21,
  author       = {Miryam Elizabeth Villa{-}P{\'{e}}rez and
                  Miguel {\'{A}}ngel {\'{A}}lvarez{-}Carmona and
                  Octavio Loyola{-}Gonz{\'{a}}lez and
                  Miguel Angel Medina{-}P{\'{e}}rez and
                  Juan Carlos Velazco{-}Rossell and
                  Kim{-}Kwang Raymond Choo},
  title        = {Semi-supervised anomaly detection algorithms: {A} comparative summary
                  and future research directions},
  journal      = {Knowl. Based Syst.},
  volume       = {218},
  pages        = {106878},
  year         = {2021},
  doi          = {10.1016/J.KNOSYS.2021.106878},
}

@inproceedings{DBLP:conf/aaai/JangK25,
  author       = {Jaeseok Jang and
                  Hyuk{-}Yoon Kwon},
  title        = {{TAIL-MIL:} Time-Aware and Instance-Learnable Multiple Instance Learning
                  for Multivariate Time Series Anomaly Detection},
  booktitle    = {Proceedings of the AAAI Conference on Artificial Intelligence},
  pages        = {17582--17589},
  publisher    = {{AAAI} Press},
  year         = {2025},
  doi          = {10.1609/AAAI.V39I17.33933},
}

@inproceedings{DBLP:conf/nips/GaoKQHTZ24,
  author       = {Shanghua Gao and
                  Teddy Koker and
                  Owen Queen and
                  Tom Hartvigsen and
                  Theodoros Tsiligkaridis and
                  Marinka Zitnik},
  title        = {UniTS: {A} Unified Multi-Task Time Series Model},
  booktitle    = {Advances in Neural Information Processing Systems 38: Annual Conference
                  on Neural Information Processing Systems (NeurIPS)},
  year         = {2024}
}

@article{zhou2023one,
  title={One fits all: Power general time series analysis by pretrained lm},
  author={Zhou, Tian and Niu, Peisong and Sun, Liang and Jin, Rong and others},
  journal={Advances in neural information processing systems},
  volume={36},
  pages={43322--43355},
  year={2023}
}

@inproceedings{DBLP:conf/icassp/LiZKQX25,
  author       = {Chengyuan Li and
                  Suyang Zhou and
                  Jieping Kong and
                  Lei Qi and
                  Hui Xue},
  title        = {KAnoCLIP: Zero-Shot Anomaly Detection through Knowledge-Driven Prompt
                  Learning and Enhanced Cross-Modal Integration},
  booktitle    = {{IEEE} International Conference on Acoustics, Speech and Signal
                  Processing (ICASSP)},
  pages        = {1--5},
  publisher    = {{IEEE}},
  year         = {2025},
  doi          = {10.1109/ICASSP49660.2025.10888834},
}

@inproceedings{DBLP:conf/iclr/ShojaeeMGFR25,
  author       = {Parshin Shojaee and
                  Kazem Meidani and
                  Shashank Gupta and
                  Amir Barati Farimani and
                  Chandan K. Reddy},
  title        = {{LLM-SR:} Scientific Equation Discovery via Programming with Large
                  Language Models},
  booktitle    = {The Thirteenth International Conference on Learning Representations,
                  (ICLR)},
  publisher    = {OpenReview.net},
  year         = {2025},
}

@article{DBLP:journals/telo/MeyersonNBGKHL24,
  author       = {Elliot Meyerson and
                  Mark J. Nelson and
                  Herbie Bradley and
                  Adam Gaier and
                  Arash Moradi Karkaj and
                  Amy K. Hoover and
                  Joel Lehman},
  title        = {Language Model Crossover: Variation through Few-Shot Prompting},
  journal      = {{ACM} Trans. Evol. Learn. Optim.},
  volume       = {4},
  number       = {4},
  pages        = {27:1--27:40},
  year         = {2024},
}

@article{DBLP:journals/nature/RomeraParedesBNBKDREWFKF24,
  author       = {Bernardino Romera{-}Paredes and
                  Mohammadamin Barekatain and
                  Alexander Novikov and
                  Matej Balog and
                  M. Pawan Kumar and
                  Emilien Dupont and
                  Francisco J. R. Ruiz and
                  Jordan S. Ellenberg and
                  Pengming Wang and
                  Omar Fawzi and
                  Pushmeet Kohli and
                  Alhussein Fawzi},
  title        = {Mathematical discoveries from program search with large language models},
  journal      = {Nat.},
  volume       = {625},
  number       = {7995},
  pages        = {468--475},
  year         = {2024}
}

@article{wang2025drsr,
  title={DrSR: LLM based Scientific Equation Discovery with Dual Reasoning from Data and Experience},
  author={Wang, Runxiang and Wang, Boxiao and Li, Kai and Zhang, Yifan and Cheng, Jian},
  journal={arXiv preprint arXiv:2506.04282},
  year={2025}
}

@article{DBLP:journals/corr/abs-2401-02981,
  author       = {Cheonsu Jeong},
  title        = {Fine-tuning and Utilization Methods of Domain-specific LLMs},
  journal      = {CoRR},
  volume       = {abs/2401.02981},
  year         = {2024}
}

@article{DBLP:journals/corr/abs-2402-15061,
  author       = {Jiawei Zheng and
                  Hanghai Hong and
                  Xiaoli Wang and
                  Jingsong Su and
                  Yonggui Liang and
                  Shikai Wu},
  title        = {Fine-tuning Large Language Models for Domain-specific Machine Translation},
  journal      = {CoRR},
  volume       = {abs/2402.15061},
  year         = {2024}
}

@inproceedings{DBLP:conf/icml/ShenTHAF24,
  author       = {Junhong Shen and
                  Neil A. Tenenholtz and
                  James Brian Hall and
                  David Alvarez{-}Melis and
                  Nicol{\`{o}} Fusi},
  title        = {Tag-LLM: Repurposing General-Purpose LLMs for Specialized Domains},
  booktitle    = {Forty-first International Conference on Machine Learning (ICML)},
  publisher    = {OpenReview.net},
  year         = {2024}
}
\end{document}